\documentclass[5p,times]{elsarticle}
\usepackage{ecrc}
\volume{00}
\firstpage{1}
\journalname{Control Engineering Practice}
\runauth{Minghao Han et al.}
\CopyrightLine{2024}{Published by Elsevier Ltd.}

\usepackage{amssymb}
\usepackage{amsmath}
\usepackage[figuresright]{rotating}
\usepackage{color}
\usepackage{subfigure}
\usepackage{arydshln}
\usepackage{algorithm}
\usepackage{algorithmic}
\usepackage{multirow} 
\usepackage{url}

\newtheorem{remark}{Remark}
\newtheorem{problem}{Problem}

\begin{document}

\begin{frontmatter}

\title{\Large \bf Efficient Economic Model Predictive Control of Water
Treatment Process with Learning-based Koopman Operator \tnoteref{abc}}
\tnotetext[abc]{This research is supported by the Ministry of Education, Singapore, under its Academic Research Fund Tier 1 (RG63/22). This research is also supported by the National Research Foundation, Singapore, and PUB, Singapore’s National Water Agency under its RIE2025 Urban Solutions and Sustainability (USS) (Water) Centre of Excellence (CoE) Programme, awarded to Nanyang Environment \& Water Research Institute (NEWRI), Nanyang Technological University, Singapore (NTU). $^1$Corresponding author: X. Yin. Tel: (+65) 6316 8746. Email: xunyuan.yin@ntu.edu.sg.
}

\author[NEWRI,CCEB]{Minghao Han}
\author[CCEB]{Jingshi Yao}
\author[NEWRI,CEE]{Adrian Wing-Keung Law}
\author[NEWRI,CCEB,1]{Xunyuan Yin}
\address[NEWRI]{Environmental Process Modelling Centre, Nanyang Environment and Water Research Institute (NEWRI), Nanyang Technological University, Singapore}
\address[CCEB]{School of Chemistry, Chemical Engineering and Biotechnology, Nanyang Technological University, Singapore}
\address[CEE]{School of Civil and Environmental Engineering, Nanyang Technological University, Singapore}

\begin{abstract}
Used water treatment plays a pivotal role in advancing environmental sustainability. Economic model predictive control holds the promise of enhancing the overall operational performance of the water treatment facilities. 
In this study, we propose a data-driven economic predictive control approach within the Koopman modeling framework. 
First, we propose a deep learning-enabled input-output Koopman modeling approach, which predicts the overall economic operational cost of the wastewater treatment process based on input data and available output measurements that are directly linked to the operational costs.
Subsequently, by leveraging this learned input-output Koopman model, a convex economic predictive control scheme is developed. The resulting predictive control problem can be efficiently solved by leveraging quadratic programming solvers, and complex non-convex optimization problems are bypassed. The proposed method is applied to a benchmark wastewater treatment process. The proposed method significantly improves the overall economic operational performance of the water treatment process. Additionally, the computational efficiency of the proposed method is significantly enhanced as compared to benchmark control solutions. 
\end{abstract}


\end{frontmatter}

\section{Introduction}

The treatment of used water is a systematic process that removes impurities from the influent feed and converts it into safer and clearer effluent that can meet specific requirements. The water treatment facilities have been playing a critical role in the preservation of the ecosystems and the protection of human health \cite{Yin_1, Yin_2}. Modern water treatment facilities typically incorporate multiple tightly interconnected physical units. Their dynamic operations involve a variety of physical, biological, and chemical phenomena, and typically exhibit complex nonlinear behaviors that are affected by significant external disturbances, including fluctuations in the flow rates and variations in the feed compositions \cite{alex2008benchmark, Yin_3}. As environmental regulations continue to tighten around the world, the industry is demanding advanced control solutions that can be used to manage the real-time dynamic operations of water treatment facilities to ensure the production of consistent high-quality treated water while maintaining the energy and chemical consumption and operating costs at a satisfactory level \cite{Yin_8, Yin_4}.

Over the past decades, research attention has been dedicated to the development of control strategies for wastewater treatment plants (WWTP). Classical proportional-integral (PI) control has been applied for ammonia removal in WWTPs \cite{vrevcko2006improvement}. 
Another commonly adopted control paradigm for WWTPs is the model predictive control (MPC). As compared to conventional control, MPC is capable of explicitly addressing constraints on quality variables and optimality considerations. In \cite{han2012model}, a self-organizing radial basis function (RBF) neural network was established as the model basis of MPC; this method can control the dissolved oxygen concentration in a WWTP. In \cite{harja2016mpc}, a hierarchical MPC design was proposed to maintain the output substrate concentration within the regulatory constraints for the biological treatment process in WWTPs. In \cite{palma2022simultaneous}, an iterative back-off approach for nonlinear model predictive control under uncertainty was proposed for the control of WWTPs. More relevant results on the MPC of water treatment processes can be found in \cite{sadeghassadi2018application,shen2008application,han2022multi,holenda2008dissolved}.
It is worth mentioning that most of the existing MPC approaches proposed for WWTPs were developed on a set-point tracking basis, that is, they aim to drive the process operation toward a specific steady-state point.
While set-point tracking MPC makes an effort to maintain the process operation around a desired stable level, it does not explicitly take into account the associated operating costs. This limitation, to some extent, restricts its capability of optimizing the overall process operation performance.
Economic model predictive control (EMPC) offers a promising alternative in the field of optimal control \cite{ellis2014tutorial,rawlings2012fundamentals,wu2022economic,liu2016economic,huang2022comparative}. By seamlessly integrating economic process optimization and dynamic control into one framework, EMPC has the potential to simultaneously optimize the effluent quality and reduce the operating costs for WWTPs \cite{rawlings2012fundamentals}. 
In \cite{Yin_8}, as one of the first attempts to address the limitations of set-point tracking MPC in the context of WWTPs, a centralized EMPC scheme was developed for a treatment process of which the dynamics are described by the benchmark simulation model no. 1 (BSM1) \cite{alex2008benchmark}. Considering the complex structure and the large scale of this process, the method in \cite{Yin_8} was extended to the distributed framework, where local EMPC controllers are deployed to coordinate their control actions through information exchange.
However, these EMPC designs face constraints in three key aspects that limit their broader applicability. First, these EMPC methods rely on an accurate first-principles model of the underlying process, which can be restrictive when accurate parameters are partially or entirely unavailable, or when the mechanistic knowledge of the nonlinear process is not completely known. Second, the implementation of these EMPC designs requires access to full-state information, which requires that all the states are measured online or a nonlinear state estimator is designed \cite{heidarinejad2012state}. Lastly, when applied to nonlinear systems, the existing EMPC designs require solving non-convex nonlinear optimization problems, which can be computationally expensive and time-consuming. This computational efficiency bottleneck may constrain the ability of EMPC to provide responsive real-time control for high-dimensional systems, such as WWTPs.

In recent years, the Koopman operator theory has gained substantial research attention, owing to its capability to represent the dynamics of complex nonlinear processes in a linear manner \cite{koopman1931hamiltonian}. The transformation of a nonlinear dynamic model into a linear model, which can also be interpreted as a global linearization paradigm of the nonlinear model, is particularly appealing since this will significantly facilitate the application of linear control theories and methods for analyzing and controlling nonlinear systems \cite{korda2018linear,proctor2018generalizing}. The success of this global linearization hinges on the quality of the Koopman operator, which governs the evolution of observables in a higher-dimensional linear state space. From a practical standpoint, determining the exact Koopman operator for a nonlinear system can be a challenging task, and it potentially involves an infinite-dimensional space. An alternative approach is to identify a well-approximated Koopman operator within a finite-dimensional space \cite{korda2018linear}. Accordingly, several cost-effective algorithms have emerged for the construction of approximated Koopman operators and Koopman linear models using temporal data of a system/process. Representative approaches include dynamic mode decomposition (DMD) \cite{schmid2010dynamic} and extended dynamic mode decomposition (EDMD) \cite{li2017extended,son2022hybrid}.
Both DMD and EDMD approximate linear Koopman operators by formulating and solving least-squares problems. As compared to DMD, its non-trivial extension, EDMD, has the potential to offer enhanced capabilities for tackling highly nonlinear processes. Meanwhile, EDMD requires the encoding of observables, typically through a manual selection of nonlinear lifting functions, which has limited the wider adoption of EDMD to some extent. This limitation arises from the considerable effort involved in the lifting function selection and the challenges of identifying appropriate functions for various industrial systems.
To streamline the design of the observable functions, researchers have leveraged deep learning and have proposed Deep-DMD Koopman modeling methods \cite{morton2018deep, han2020deep,shi2022deep}. Building upon this foundation, a Koopman autoencoder framework was proposed for learning Koopman models \cite{azencot2020forecasting}. 
More results on Koopman-based modeling and control can be referred to \textcolor{black}{\cite{korda2018power, cibulka2020model, yeung2019learning,bruder2019modeling,han2021desko,son2021application,zhang2023reduced,han2023robust_tii,han2023robust_tnnls,narasingam2023data,son2022development,narasingam2019koopman,son2022development}}.

In addition to the Koopman operator methods, hybrid modeling has been proposed to seamlessly integrate first-principles knowledge with data information, aiming at enhancing data efficiency and accuracy, and much progress has been achieved in recent years \cite{bangi2020deep,shah2022deep}. \cite{bangi2020deep,shah2022deep} combined deep learning with first-principles models to capture the dynamics of chemical processes, and achieved higher accuracy with less data. Another promising framework is the physics-informed machine learning (PIML). According to \cite{karniadakis2021physics}, PIML holds the potential to reduce the amount of data and1 yet can provide improved accuracy and generalization. In recent years, exciting progress has been made \cite{raissi2019physics,sirignano2020dpm,lu2020extraction} with diverse applications to quantum chemistry \cite{pfau2020ab}, material sciences \cite{shukla2020physics}, molecular simulations \cite{zhang2018deep}, etc. 

It has been recognized that the Koopman operator provides a promising alternative approach to address complex nonlinear control problems efficiently by leveraging linear control theories, such that the computational complexity can be substantially reduced. This motivates us to explore whether the Koopman modeling framework may be exploited for efficient EMPC. This is of particular interest because, while EMPC can be more adaptable than set-point tracking MPC and can be more favourable for nonlinear industrial processes like WWTP from an economic and sustainable operation point of view,
the potential of this important variant of MPC has been hindered by excessive computational expenses caused by the use of typically nonlinear objective functions and nonlinear dynamic process models.

In this work, we aim to bridge this gap by proposing a Koopman-based economic predictive control framework, which is designed to alleviate the prohibitive computational burden associated with the existing non-convex EMPC frameworks to facilitate the efficient and economic operation of WWTPs. Specifically, we propose a data-driven modeling framework, referred to as the Deep Input-output Koopman Operator (DIOKO) model, to characterize the relationship between the control inputs and the overall economic operational cost. By using only input data and economic cost data, the proposed modeling approach learns a computationally efficient latent space, where future key performance indices related to WWTP operations are accurately predicted. Based on the established DIOKO model, a convex computationally efficient EMPC scheme is formulated. This data-driven EMPC control scheme is applied for the real-time operation of the nonlinear WWTP, which leads to significantly improved overall operational performance under all three weather conditions. The contributions of this work are threefold:
\begin{itemize}
    \item We propose a learning-based input-output Koopman modeling approach to predict future economic performance indices of the WWTP without requiring full-state information of the process.
    \item A data-driven convex economic predictive control scheme is developed for the economic operation of the WWTP which exhibits highly nonlinear dynamics.
    \item As compared to conventional EMPC based on a nonlinear first-principles model, our proposed framework achieves an over 5600-fold improvement in computational efficiency, while simultaneously delivering overall better economic operational performance.
\end{itemize}
Some preliminary findings from this research were previously submitted in a conference paper \cite{minghao_acc}. As compared to the conference version of this work \cite{minghao_acc}, the current paper presents more comprehensive descriptions and an in-depth exploration of the proposed framework. In the conference version \cite{minghao_acc}, we conducted simulations to showcase the effectiveness of the proposed method under the dry weather condition. In the current paper, we substantially expand our simulations and present more extensive results and comparative analysis to illustrate the advantages of the proposed framework under all three major weather conditions, including dry, rainy, and stormy weather conditions. Additionally, in this paper, we include explanatory remarks to elucidate the rationale behind the design of the economic output function adopted in the proposed data-driven EMPC scheme, and to discuss the advantages of the proposed method over the existing EMPC methods. Last but not least, we also evaluate the generalizability of the proposed approach to scenarios that were not encountered during the training phase. This assessment aids in analyzing the reliability and the broader applicability of the proposed framework.

The remainder is structured as follows. In Chapter 2, we provide a comprehensive overview of the wastewater treatment plant and a description of the key performance indices that are associated with the operational efficiency and cost of the plant. In Chapter 3, the data-driven economic model predictive control problem is formulated. Chapter 4 presents the proposed control method.  Chapter 5 presents the results obtained based on the proposed method, and compares the performance with a benchmark EMPC design. Conclusions are drawn in Section 6. 
\section{ System description and performance indices}

In this section, we introduce the wastewater treatment plant (WWTP) and describe the key performance indices that are used to assess the effectiveness and performance of the water treatment process operations. 

\subsection{System description}\label{section:2:1}
\begin{figure}
    \centering
    \includegraphics[width=0.5\textwidth]{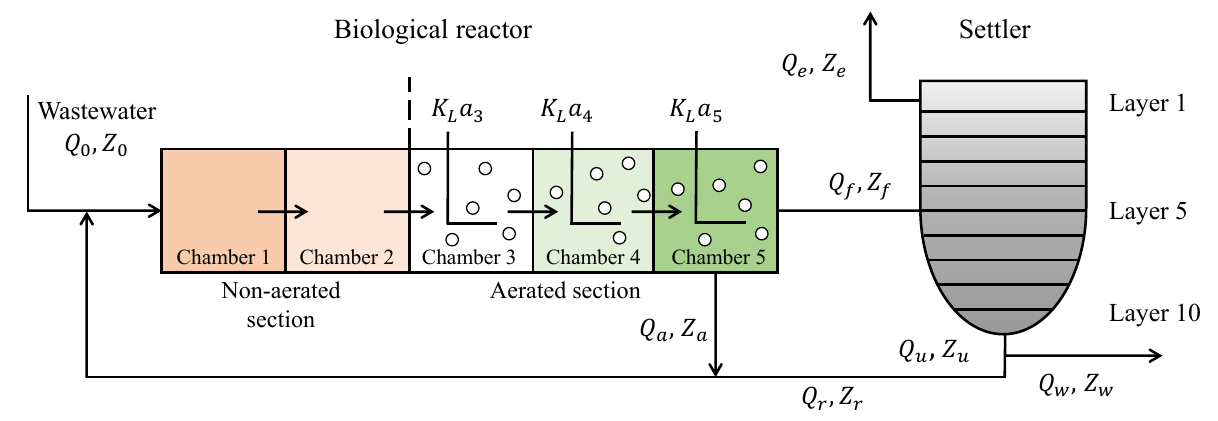}
    \caption{A schematic of the biological wastewater treatment plant \cite{alex2008benchmark}.}
    \label{fig:plant_overview}
\end{figure}

In this work, we consider a benchmark WWTP, of which the dynamics are described by the benchmark simulation model no.1 (BSM1) \cite{alex2008benchmark}. This WWTP consists of a five-compartment activated sludge reactor and a secondary settler, as shown in Figure~\ref{fig:plant_overview}. The five compartments are: the first two are anoxic compartments which form the non-aerated section, and the subsequent three are aerobic compartments which form the aerated section. Wastewater is fed into the first compartment of the activated sludge reactor at a flow rate of $Q_\text{0}$, with a concentration of $Z_0$. 
Denitrification biological processes take place within the two anoxic compartments; in which the bacteria converts the nitrate into nitrogen. At the same time, nitrification biological processes take place in the three aerobic compartments, in which the bacteria converts the ammonium into nitrate through oxidation. 
A return stream, originating from the last aerated compartment, is redirected back into the first anoxic compartment at a flow rate of $Q_\text{a}$ with a concentration of $Z_\text{a}$. The remainder of the effluent is fed into the secondary settler at flow rate $Q_\text{f}$ and concentration $Z_\text{f}$.
The secondary settler has 10 non-reactive layers, with the fifth layer (counting from the top to the bottom) being the feed layer.
A portion of the substance present in the settler is reintroduced into the first anoxic compartment through an external recycle stream at a flow rate of $Q_\text{r}$ and a concentration of $Z_\text{r}$. 
Concurrently, excess sludge is removed from the bottom layer of the secondary settler at flow rate $Q_\text{w}$ and concentration $Z_\text{w}$. Processed water is extracted from the top layer of the settler at flow rate $Q_\text{e}$ and concentration $Z_\text{e}$.

\begin{table}[tb]
\centering
\caption{A list of the 8 biological reactions taking place in the activated sludge reactor.}\vspace{2mm}
\label{table:Biological Processes}
\renewcommand\arraystretch{1.27}
\begin{tabular}{c c}
\hline
Index & Reactions\\
\hline
1& aerobic growth of heterotrophs\\
2& anoxic growth of heterotrophs\\
3& aerobic growth of autotrophs\\
4& decay of heterotrophs\\
5& decay of autotrophs\\
6& ammonification of soluble organic nitrogen\\
7& hydrolysis of entrapped organics\\
8& hydrolysis of entrapped organic nitrogen\\
\hline
\end{tabular}
\end{table}

In this plant, 8 biological reactions take place in the activated sludge reactor, and they are summarized in Table 1. 13 major compounds that are involved in the 8 reactions are taken into account. The concentrations of the 13 compounds in each compartment of the activated sludge reactor are the state variables of the corresponding compartment. The state variables in each of the compartments of the reactor are defined in Table 2. The dynamical behaviors of the 13 state variables of each compartment are characterized by 13 ordinary differential equations (ODEs). 
In each layer of the 10-layer secondary settler, we consider 8 state variables, including $S_{\text{I}}$, $S_{\text{S}}$, $S_{\text{O}}$, $S_{\text{NO}}$, $S_{\text{NH}}$, $S_{\text{ND}}$, $S_{\text{ALK}}$ which are defined in Table 1, and $X$ which represents the concentration of the total sludge (a weighted summation of $X_{\text{I}}$, $X_{\text{S}}$
, $X_{\text{B,H}}$
, $X_{\text{B,A}}$
, $X_{\text{P}}$
, and $X_{\text{ND}}$). 
The concentration profile of the solids in the settler is analyzed by conducting a solids balance calculation for each layer.
The dynamics of state variables of the settler are characterized by ODEs that are established based on the mass balances of the sludge, assuming that the particulate concentrations entering the settler can instantaneously disperse to both the upper and lower layers of the settler, that is, the sludge retention time within the settler is disregarded. Therefore, the dynamic model for the entire plant consists of 145 ODEs -- 65 ODEs governing the dynamical behaviors of the reactor and 80 ODEs for the settler \cite{alex2008benchmark,Yin_8}.
A comprehensive description of the benchmark simulation model is referred to \cite{alex2008benchmark}. The values of the key parameters of this plant are presented in Table 3. For this process, we consider two manipulated input variables: the internal recycle flow rate denoted by $Q_\text{a}$, and the oxygen transfer coefficient in the fifth compartment of the reactor, which is denoted by $K_\text{L}a_5$.

\begin{table}[tb]
\centering
\caption{A list of the 13 state variables of the activated sludge reactor.}\vspace{2mm}
\label{table:state variables}\renewcommand\arraystretch{1.27}
\begin{tabular}{p{1.1cm}|p{4.5cm}|c}
\hline
Notation&Definition & Unit\\
\hline
$S_\text{I}$& soluble inert organic matter &$\text{gCOD}/\text{m}^3$\\
$S_\text{S}$& readily biodegradable substrate &$\text{gCOD}/\text{m}^3$\\
$X_\text{I}$& particulate inert organic matter &$\text{gCOD}/\text{m}^3$\\
$X_\text{S}$& slowly biodegradable substrate &$\text{gCOD}/\text{m}^3$\\
$X_\text{B,H}$& active heterotrophic biomass &$\text{gCOD}/\text{m}^3$\\
$X_\text{B,A}$& active autotrophic biomass &$\text{gCOD}/\text{m}^3$\\
$X_\text{P}$& particulate products arising from biomass decay &$\text{gCOD}/\text{m}^3$\\
$S_\text{O}$& oxygen &$\text{g(-COD)}/\text{m}^3$\\
$S_\text{NO}$& nitrate and nitrite nitrogen &$\text{gN}/\text{m}^3$\\
$S_\text{NH}$& $\text{NH}_4^+$ and $\text{NH}_3$ nitrogen &$\text{gN}/\text{m}^3$\\
$S_\text{ND}$& soluble biodegradable organic nitrogen &$\text{gN}/\text{m}^3$\\
$X_\text{ND}$& particulate biodegradable organic nitrogen &$\text{gN}/\text{m}^3$\\
$S_\text{ALK}$& alkalinity &$\text{mol}/\text{m}^3$\\
\hline
\end{tabular}
\end{table}

\begin{table}[tb]
\centering
\caption{Parameters of BSM1}\vspace{2mm}
\label{table:Model Parameters}\renewcommand\arraystretch{1.27}
\begin{tabular}{l l}
\hline
Definitions and notations & Unit\\
\hline
volume of compartment 1 ($V_1$) & 1000 $\text{m}^3$ \\
volume of compartment 2 ($V_2$) & 1000 $\text{m}^3$ \\
volume of compartment 3 ($V_3$) & 1333 $\text{m}^3$ \\
volume of compartment 4 ($V_4$) & 1333 $\text{m}^3$ \\
volume of compartment 5 ($V_5$) & 1333 $\text{m}^3$ \\

wastage flow rate ($Q_\text{w}$) & 385 $\text{m}^3/\text{day}$ \\
volume of settler & 6000 $\text{m}^3$\\
cross-sectional area of settler ($A$) & 1500 $\text{m}^2$\\
height of settler & 4$\text{m}$ \\
height of each layer ($H_j, j = 1, \dots, 10$) & 0.4 $\text{m}$\\
number of layers & 10 \\
return sludge flow rate ($Q_\text{r}$) & 18846 $\text{m}^3/\text{day}$\\
\hline
\end{tabular}

\end{table}

\subsection{ Performance indices}\label{sec:performance_indices}

In this work, a few key performance indices are used to verify the effectiveness and assess the performance of process operations governed by different control designs. 
These indices are used to assess various aspects of the process operation performance, including the quality of the effluent discharged from the settler, the overall operating cost, and the set-point tracking performance. 

The effluent quality, denoted by EQ (kg  pollutant/day),
which is concerned with the effluent discharged from the top layer of the secondary settler, and is a key indicator of the performance of the operation of the WWTP \textcolor{black}{being representative of the environmental impact of the treated water discharged from the water treatment plant.}
EQ is defined as the daily average computed from a weighted summation of the concentrations of various compounds contained in the effluent over a specified time period. Specifically, EQ is computed as follows \cite{alex2008benchmark}: 
{\small
\begin{align}
    \text{EQ} = \frac{1}{1000T} \int_{t_0}^{t_0+T}  \biggr( &a_1\text{TSS}_\text{e}\left(t\right)  
    +a_2\text{COD}_\text{e}\left(t\right)
    +a_3\text{S}_\text{NKj,e}\left(t\right)\notag\\
    &+a_4\text{S}_\text{NO,e}\left(t\right)
    +a_5\text{BOD}_\text{e}\left(t\right)
     \biggr)\text{Q}_\text{e}\left(t\right) \text{d}t.\label{eq:EQ}
\end{align}}
In \eqref{eq:EQ}, $t_0$ is the initial time instant; $T$ denotes the overall operational duration; $\text{TSS}$ denotes the total suspended solids; $\text{COD}$ represents the chemical oxygen demand; $\text{S}_\text{NKj}$ denotes the Kjeldahl nitrogen concentration; $\text{BOD}$ represents the biological oxygen demand. The subscript “e” in any of the symbols involved in (1) indicates that these variables pertain to the effluent of the settler. In this study, the weights $a_i, i=1, \dots, 5$ for $\text{TSS}_\text{e}$, $\text{COD}_\text{e}$, $\text{S}_\text{NKj,e}$, $\text{S}_\text{NO,e}$, and $\text{BOD}_\text{e}$, respectively, are made the same as those in \cite{Yin_8,vanrolleghem1996integration}, i.e. $a_1 = 2$, $a_2 = 1$, $a_3 = 30$, $a_4 = 10$, $a_5 = 2$.
For a more in-depth understanding of these variables, please refer to \cite{jeppsson2004cost} for detailed expressions. These weights can also be further tuned to suit a specific treatment facility.

The operating cost is also an important performance indicator. The operating cost comprises the amount of produced sludge to be disposed of, the aeration energy, the pumping energy, and the mixing energy. The sludge
production, denoted by $\text{SP}$ ($\text{kg}/\text{day}$), represents the daily average of the total solids in the wastage flow (i.e. $\text{Q}_\text{w}$) and solids accumulated in the plant during the overall operational duration $T$:
\begin{equation}
    \begin{aligned}
        \text{SP} = &\frac{1}{1000T} (0.75\int_{t_0}^{t_0+T} \left( \text{X}_\text{s,w}\left(t\right)  
    +\text{X}_\text{I,w}\left(t\right)
    +\text{X}_\text{B,H,w}\left(t\right)\right. \\
    &\left.\left. +\text{X}_\text{B,A,w}\left(t\right)
    +\text{X}_\text{P,w}\left(t\right)
    \right) \right)\text{Q}_\text{w}\left(t\right) \text{d}t + \left( \text{TSS}\left(t_0+T\right) - \text{TSS}\left(t_0\right) \right)
\end{aligned}
\end{equation}
where $\text{TSS}(t)$ is the amount of solids in the plant at the time instant $t$. Details of the calculation of $\text{TSS}$ are referred to \cite{alex2008benchmark}.

The aeration energy $\text{AE}$ ($\text{kWh}/\text{day}$), which refers to the energy needed for the process of aeration, is determined by the design of the facility, including the type of diffuser, the size of the bubbles, and the depth of the submersion. According to \cite{alex2008benchmark}, AE can be calculated based on the oxygen transfer coefficients as follows:
\begin{equation}
    \text{AE} = \frac{S_\text{o}^\text{sat}}{1800T}\int_{t_0}^{t_0+T} \sum^5_{i=1} V_iK_\text{L}a_i(t) \text{d}t
\end{equation}
where $S_\text{o}^\text{sat}$ represents the saturation concentration for oxygen
of which the value is set to $8\text{g}/\text{m}^3$ in this study; $K_\text{L}a_i$ denotes the oxygen transfer coefficient in the $i$th compartment of the activated sludge reactor, $i=1,\ldots,5$.

The pumping energy \text{PE} ($\text{kWh}/\text{day}$) accounts for the energy consumed by the pumps in both internal and external recycling processes. It is determined as:
\begin{equation}
    \text{PE} = \frac{1}{T}\int_{t_0}^{t_0+T} (
    b_1\text{Q}_\text{a}(t) 
    +b_2\text{Q}_\text{r}(t) 
    + b_3\text{Q}_\text{w}(t) ) \text{d}t
\end{equation}
where $b_1=0.004$, $b_2=0.008$, $b_3=0.05$, according to \cite{alex2008benchmark}.

The mixing energy, ME ($\text{kWh}/\text{day}$), which refers to the energy consumption related to mixing the anoxic state to avoid settling in the compartments of the reactor, is dependent on the volume of the reactor compartments and the oxygen transfer coefficient inside the compartments. It is computed as follows \cite{alex2008benchmark}:
\begin{equation}
    \text{ME} = \frac{1}{T}\int_{t_0}^{t_0+T} \text{ME}_f(t) \text{d}t
\end{equation}
where
\begin{equation}
    \text{ME}_f(t) =24\sum^5_{i=1} \left\{
    \begin{matrix}
m_1V_i &\text{  if }K_\text{L}a_i(t)<20\text{/day} \\
0 &\text{  if }K_\text{L}a_i(t)\geq20\text{/day}
\end{matrix}\right.
\end{equation}
where $m_1=0.005$.

The overall cost index, denoted by $\text{OCI}$, is assessed by jointly taking into account the sludge production, aeration energy, pumping energy, and mixing energy using the following equation \cite{alex2008benchmark, Yin_8}:
\begin{equation}
    \text{OCI} = 5\text{SP} +\text{AE} + \text{PE} + \text{ME}.
\end{equation}

The computation of indices of EQ and OCI is dependent on 41 state variables out of all the 145 state variables of the entire plant.
To enhance the effluent quality produced by the WWTP while improving the overall process operational efficiency and performance, it is desirable to regulate the real-time operation of the WWTP in a way such that both the effluent quality index EQ and the overall operating cost index OCI are simultaneously reduced.

\section{Problem formulation}

The dynamics of the considered WWTP can be described by the following general compact nonlinear state-space form:
\begin{subequations}
\begin{align}
     x_{k+1}  &= f(x_k, u_k, d_k) \label{eq:process model:f}\\
     y_k &= h(x_k)  \label{eq:process model:h}\\
     c_{k} &= w_\text{EQ}\text{EQ}_k + w_\text{OCI}\text{OCI}_k \label{eq:stage cost}
\end{align}\label{eq:process model}
\end{subequations}
\vspace{-0.6cm}

\noindent
where $k\in \mathbb{N}$ denotes the time instant; 
$x  \in \mathbb{R}^{145}$ denotes the state vector; 
$y  \in \mathbb{R}^{41}$ denotes the vector of sensor measurements; 
$c  \in \mathbb{R}$ denotes the overall economic stage cost;
$u \in \mathbb{R}^m$ denotes the control input vector; 
$d  \in \mathbb{R}^p$ denotes the vector of known disturbances, which include the flow rate of the wastewater entering the first compartment of the bioreactor and the concentrations of the 13 compounds in the inlet wastewater flow. In (\ref{eq:process model}),
$f(\cdot)$ is a nonlinear function characterizing the state dynamics of the process; $h(\cdot)$ denotes the output measurement functions; $c_k$ in \eqref{eq:stage cost}, referred to as the overall economic stage cost associated with the process operation, summarizes the key performance indices described in Section~\ref{sec:performance_indices} for the time instant $k$. $w_\text{EQ}$ and $w_\text{OCI}$ are the weighting coefficients for the effluent quality and overall cost index, respectively. They can be adjusted to achieve a balance between the importance placed on treated water quality and economic costs. The vector of sensor measurements has 41 elements, which are the state variables out of all the 145 state variables needed for the calculation of EQ and OCI. 

Assuming that the nonlinear state-space model in (\ref{eq:process model:f}) is available, and that all the 145 state variables are measured online using sensors, then the optimization problem for a conventional first-principles model-based EMPC design can be formulated as Problem~\ref{problem:1} below:

\begin{problem}\label{problem:1} \cite{liu2015economic}
    Given the nonlinear process model of the WWTP in the form of (\ref{eq:process model:f}), full-state measurements at each sampling instant $k$, the objective of the EMPC is to determine a sequence of optimal control inputs $u^*_{k:k+T_\text{f}|k}=\small\{u^*_{k|k}, u^*_{k+1|k},\ldots, u^*_{k+T_\text{f}-1|k}\small\}$ that minimizes the sum of the stage costs over a time horizon $T_\text{f}$, described as follows:
    \begin{subequations}\label{obj:problem1}
    \begin{align}
        \min_{u_{k:k+T_{\text{f}}|k}} \quad\sum_{j=k}^{k+T_{\text{f}}}& c_j  \\
        \text{s.t.} \quad { x}_{j+1|k}  &= f({x}_{j|k}, u_{j|k}, d_j)\\
       { x}_{k|k} &= x_k\\
         x_j &\in \mathbb{X} \\
         u_j &\in \mathbb{U}
    \end{align}
    \end{subequations}
where $u_{j|k}$ is the optimal control input for time instant $j$, $j=k,\ldots,k+T_\text{f}$, obtained by solving (\ref{obj:problem1}) at time instant $k$.
\end{problem}

However, as has been discussed in the introduction, this type of EMPC design may encounter three major limitations from an application perspective. 
\begin{itemize}
    \item This method requires the availability of the first-principles model in (\ref{eq:process model:f}) with accurate model parameters, which may be difficult to obtain.
    \item  It also requires access to the full-state information $x_k$ at each sampling instant. However, the deployment of hardware sensors to measure them comprehensively will be challenging.
    \item The formulated nonlinear programming problem is non-convex, and can be highly computationally expensive. The complex and expensive computation may restrict the capability of this design in terms of achieving real-time economic control of the WWTP of high dimensionality.
\end{itemize}

In this work, we aim to address the three limitations by developing a convex thus computationally efficient EMPC control approach for real-time optimal control of the WWTP based on process data, bypassing the need for a nonlinear first-principles model in the form of (\ref{eq:process model:f}). Specifically, the economic control objective can be described by Problem~\ref{problem:2} as follows:

\begin{problem}\label{problem:2}
    For the wastewater treatment process described in Section~\ref{section:2:1}, the objective is to propose a data-driven convex economic predictive control design to determine a sequence of optimal control input $u^*_{k:k+T_\text{f}|k}=\small\{u^*_{k|k}, u^*_{k+1|k},\ldots, u^*_{k+T_\text{f}-1|k}\small\}$ that minimizes the sum of the inferred stage costs over a time horizon $T_{f}$, described as follows:
    \begin{subequations}
    \begin{align}
        \min_{u_{k:k+T_\text{f}|k}}& \quad\sum_{j=k}^{k+T_{\text{f}}} \hat{c}_\theta ({ \psi}_{j|k}) \\
        \text{s.t.} &\quad \psi_{j+1|k}  = A\psi_{j|k} + Bu_{j|k}\\
            &\quad \psi_{k|k} = \psi_\theta(y_k, d_k)\\
            &\quad u_k \in \mathbb{U}
    \end{align}
    \end{subequations}
where $\psi_\theta(\cdot)$ is a nonlinear function of the output measurements and known disturbances with $\theta$ being parameters to be identified/learned from data; 
$\hat{c}_\theta(\cdot)$ is an approximated function of the true stage cost function $c$, and can be expressed as a quadratic function of observable vectors $\psi_{j|k}$, $j=k\ldots,k+T_{\text{f}}$; $A$ and $B$ constant matrices to characterize the dynamics of the encoded latent state $\psi_k$. The trainable parameters $\theta$, $A$, and $B$ that will be learned from a data set comprising input data $u_k$ and $d_k$, and sensor measurements $y_k$, $k\in\mathbb N$. This way, the establishment of a learned linear model as the foundation for data-driven EMPC, along with the inclusion of a quadratic approximation for the overall economic stage cost, leads to the formulation of a convex quadratic programming problem.
\end{problem}

\section{Proposed data-driven economic predictive control approach}

In this section, we propose a data-driven economic predictive control approach that will be used to control the operation of the nonlinear WWTP in a computationally efficient manner. First, we briefly review the concept of Koopman operator. Based on the Koopman operator framework, we develop an input-output Koopman modeling approach to learn a linear model using the data provided by the available sensors for the 41 measured process states involved in (\ref{eq:process model:h}). The learned model will be used to infer the overall economic stage cost $c_k$ using a quadratic form, despite the nonlinear dependence of $c_k$ in the process states. Then, based on the learned input-output Koopman model, a convex economic model predictive control system is developed for the optimal operation of the WWTP.

\begin{figure*}
    \centering
    \includegraphics[width=\textwidth]{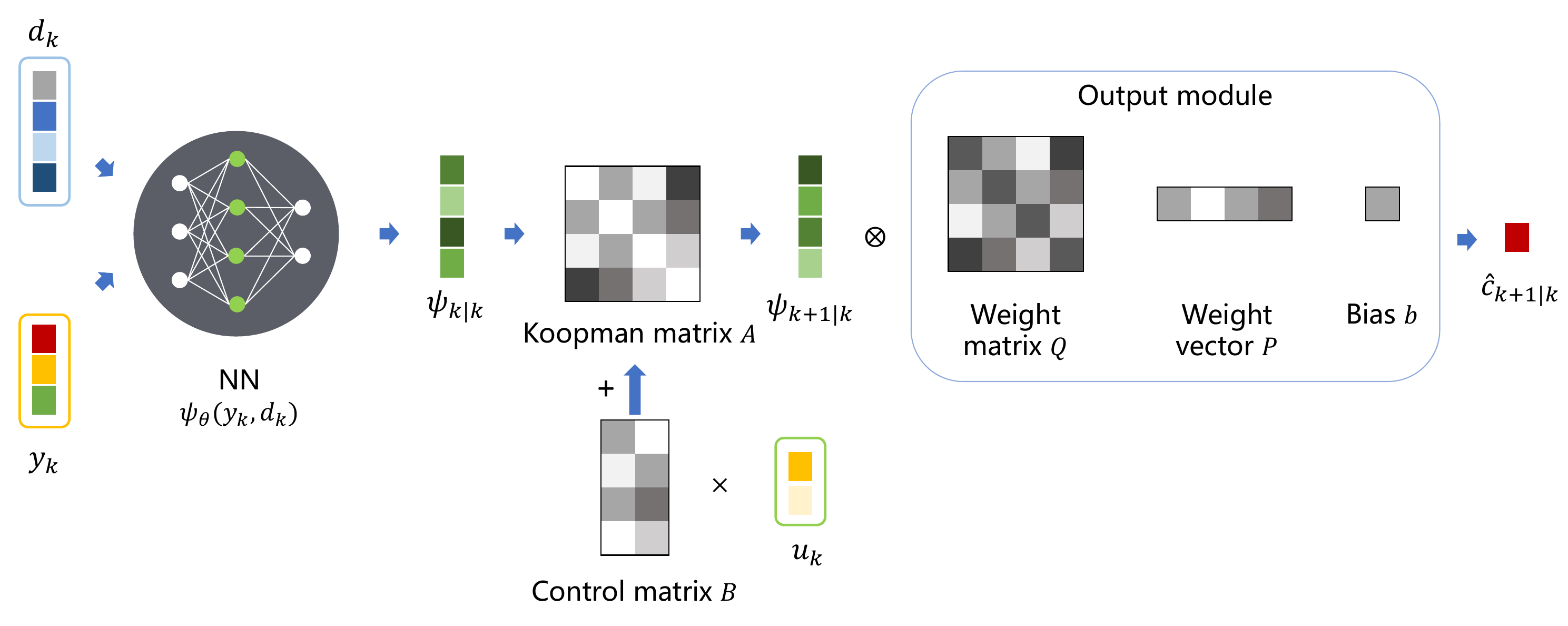}
    \caption{The pipeline of the proposed DIOKO modeling approach. The neural network encodes the measurement vector $y_k$ and the known disturbance $d_k$ into the observable vector $\psi_{k|k}$. Then, the learned Koopman matrix $A$ and matrix $B$ propagate $\psi_{k|k}$ and $u_k$ to the next step, and produce observable vector $\psi_{k+1|k}$. 
    Finally, the predicted observable vector $\psi_{k+1|k}$ is transformed by the output module through equation \eqref{eq:output function} 
    to predict the next step stage cost $\hat{c}_{k+1|k}$.}
    \label{fig:overview_koopman}
\end{figure*}

\subsection{Basis of Koopman operator}

The Koopman operator theory was initially proposed in \cite{koopman1931hamiltonian}. A Koopman operator is a higher-dimensional linear operator that operates on nonlinear functions of the state variables of the original nonlinear systems. This way, the dynamics of a nonlinear system may be transformed into linear dynamics within a higher-dimensional linear state space.

Specifically, let $x_k\in \mathbb{R}^n$ denote the state vector of a nonlinear autonomous system of which the dynamics are described by the following state-space form:
\begin{equation}\label{eq:general:nonlinear}
    x_{k+1}=f(x_k)
\end{equation}
According to Koopman theory \cite{koopman1931hamiltonian}, for the nonlinear system (\ref{eq:general:nonlinear}), there exists a higher-dimensional (up to infinite-dimensional) observable function space, denoted by $\mathcal{G}$. This space comprises all square-integrable real-valued functions defined on the compact domain $\mathbb{X}\subset\mathbb{R}^{n}$. Within this higher-dimensional state-space, the Koopman operator $\mathcal{K}:\mathcal{G} \rightarrow \mathcal{G}$ governs the dynamics of the observables in a linear manner, as follows:
\begin{equation}
    \psi(f(x_k)) = \mathcal{K}\psi(x_k),
\end{equation}
where $\psi\in \mathcal{G}$ is an \textit{observable functions}. 

From a practical perspective, problems defined within infinite-dimensional function spaces need to be avoided. Based on this consideration, let $\overline{\mathcal{G}}\subset \mathcal{G}$ denote the space spanned by a set of linearly independent observables $\{\psi|\psi:\mathbb{R}^{n}\rightarrow\mathbb{R}^p\}$, where the dimensionality $p$ is a finite hyper-parameter that can be specified by users. Then, instead of seeking the exact Koopman operator $\mathcal{K}$, a more practical approach is to identify a finite-dimensional Koopman operator $\overline{\mathcal{K}}$. This finite-dimensional Koopman operator can effectively approximate the dynamics of the underlying nonlinear system within the $\overline{\mathcal{G}}$ subspace.

While the Koopman operator was originally proposed for autonomous nonlinear systems, its application has been extended to controlled systems \cite{proctor2018generalizing}.
 Consider a nonlinear controlled system described by the following general nonlinear state-space form:
\begin{equation}
    x_{k+1} = f(x_k, u_k)
\end{equation}
where $u_k\in \mathbb{U}\subset \mathbb{R}^m$ denote the input vector at time instant $k$.

 When considering the system input $u$ in the Koopman operator, the original state is extended from $x_k$ to $\chi_k=[x_k^T, u_k^T]^T$ and the mapped state is extended from $\psi(x_k)$ to $\psi_\chi(\chi_k)$ \cite{korda2018linear}:
\begin{equation}
    \psi_\chi(\chi_k) = \left[ \begin{matrix}
        \psi_x(x_k) \\
        \psi_u(u_k)
    \end{matrix}\right]
\end{equation}
and the new Koopman operation can be represented as:
\begin{equation}
    \psi_\chi(\chi_{k+1}) = \mathcal{K}\psi_\chi(\chi_k) 
\end{equation}

Further, the finite-dimensional Koopman operator $\overline{\mathcal{K}}$ can be represented by a block matrix as:
\begin{equation}
    \overline{\mathcal{K}} = \left[ \begin{array}{c:c}
        A & B \\
        \hdashline
        * & *
    \end{array}\right]
\end{equation}

Given our primary concern for the variable $\psi(x_k)$ instead of $\psi_\chi(\chi_k)$, it is sufficient to determine matrices $A$ and $B$, enabling to establish the following Koopman model:
\begin{equation}
\psi_x(x_{k+1}) = A \psi_x(x_k) +  B \psi_u(u_k) \label{eq:deterministic Koopman with control}
\end{equation}
where $A\in \mathbb{R}^{p\times p}$; $B\in \mathbb{R}^{p\times r}$;
$\psi_x:\mathbb{R}^{n}\rightarrow\mathbb{R}^p$ and $\psi_u:\mathbb{R}^{m}\rightarrow\mathbb{R}^r$ are observables of the state and inputs, respectively.

\subsection{Deep input-output Koopman operator}
As described previously, measuring the full system state is a challenging task. In this work, instead of learning the Koopman operator and developing a control method that requires full state information, we propose a data-driven modeling approach, referred to as the deep input-output Koopman operator, to learn a model that aims to predict the future overall economic stage cost $c_k$ by utilizing the available measurements from the 41 state sensors, the control inputs, and known disturbances of the WWTP in (\ref{eq:process model}). 

An overview of the pipeline for the proposed method is presented in Figure~\ref{fig:overview_koopman}.
First, a multi-layer neural network $\psi_\theta(\cdot)$ with trainable parameters $\theta$ is employed to encode the observables $\psi_k$.
Specifically, the output measurement $y_k$ and the known disturbance to the plant $d_k$ are fed into the neural network to encode an informative observable $\psi_{k|k}$, where
\begin{equation}\label{eq:initial value}
    \psi_{k|k} = \psi_\theta(y_k, d_k).
\end{equation}

\begin{table*}
\centering
\caption{Initial Condition of the States in the Biological Reactor}\vspace{2mm}
\label{table:initial condition of BR}\renewcommand\arraystretch{1.27}
\begin{tabular}{p{2.2cm}|p{2.1cm}p{2.1cm}p{1.9cm}p{1.9cm}p{1.9cm}|p{1.9cm}}
\hline
Compartment index $i$ & $1$ &  $2$ & $3$ & $4$ &$5$  & units\\
\hline
$S_\text{I,i}$& $30$ & $30$ &$30$ &$30$ &$30$ &$\text{gCOD}/\text{m}^3$\\
$S_\text{S,i}$& $3.24$ & $1.67$ & 1.22 & 0.97 & 0.81 &$\text{gCOD}/\text{m}^3$\\
$X_\text{I,i}$& 1149.21 & 1149.21 & 1149.21 & 1149.21& 1149.21 &$\text{gCOD}/\text{m}^3$\\
$X_\text{S,i}$& 98.60 & 91.70 & 69.69 & 54.45 & 44.48 &$\text{gCOD}/\text{m}^3$\\
$X_\text{B,H,i}$& 2552.12 & 2552.39 & 2560.22 & 2563.33 & 2562.87 &$\text{gCOD}/\text{m}^3$\\
$X_\text{B,A,i}$& 151.67& 151.53& 152.69& 153.71& 154.17 &$\text{gCOD}/\text{m}^3$\\
$X_\text{P,i}$& 446.96& 448.12& 449.67& 451.22& 452.77 &$\text{gCOD}/\text{m}^3$\\
$S_\text{O,i}$& $7.696\times 10^{-3}$& $6.027\times 10^{-5}$& 1.63& 2.47& 2.00 &$\text{g(-COD)}/\text{m}^3$\\
$S_\text{NO,i}$& 3.51& 1.00& 6.23& 11.07& 13.52 &$\text{gN}/\text{m}^3$\\
$S_\text{NH,i}$& 11.83& 12.55& 7.32& 2.78& 0.67 &$\text{gN}/\text{m}^3$\\
$S_\text{ND,i}$& 1.36& 0.79& 0.83& 0.75& 0.66 &$\text{gN}/\text{m}^3$\\
$X_\text{ND,i}$& 6.18& 5.95& 4.71& 3.84& 3.26 &$\text{gN}/\text{m}^3$\\
$S_\text{ALK,i}$& 5.34& 5.57& 4.82& 4.15& 3.83 &$\text{mol}/\text{m}^3$\\
\hline
\end{tabular}
\end{table*}

The encoded observable vector $\psi_{k|k}$ is then propagated with Koopman operators $A$ and $B$ to produce the predicted next-step observable vector $\psi_{k+1|k}$, through the linear state-space form below:
\begin{equation}\label{equation:linear_model:Koopman}
    \psi_{k+1|k} = A \psi_{k|k} + Bu_k.
\end{equation}
Note that in (\ref{equation:linear_model:Koopman}), the observables of the inputs are chosen to be the control $u_k$ itself, which is a special case of the general form in \eqref{eq:deterministic Koopman with control}. This special form preserves the convexity of the resulting optimal control problem and helps improve the computational efficiency of the resulting control method. In our simulations, we observe that this parameterization yields excellent modeling performance, and we will present the simulation results in the following section. 

Finally, the predicted observable vector is passed through an output function to estimate the next-step cost, denoted by $\hat{c}_{k+1|k}$, using the following formula:
\begin{equation}\label{eq:output function}
    \hat{c}_{k+1|k} = \psi_{k+1|k}^\text{T} Q \psi_{k+1|k} + P\psi_{k+1|k} + b.
\end{equation}
In (\ref{eq:output function}), $Q = \text{diag}\small\{\exp(q_\text{v})\small\}$ and $q_\text{v}\in \mathbb{R}^{p}$ is a real-valued vector, where $\exp(\cdot)$ denotes an operator that takes the element-wise exponential of the given vector, and $\text{diag}\small\{\cdot\small\}$ signifies the construction of a diagonal matrix with the given vector as its diagonal elements. Additionally, $P\in\mathbb{R}^{1\times p}$ is a real-valued vector, and $b$ represents a constant bias.
The parameterization in \eqref{eq:output function} facilitates the formulation of the economic control problem as a quadratic programming problem, which is the key factor that will enable efficient optimization-based control of the nonlinear WWTP based on the proposed approach. This output function in the form of \eqref{eq:output function} is only one of the possible parameterized approximation realizations for mapping the observable vector to the overall economic stage cost $c$.

In the proposed pipeline, $\theta$, $A$, $B$, $q_\text{v}$, $P$, and $b$ are trainable parameters, which are updated by minimizing the following finite-horizon objective function:
\begin{subequations}\label{eq:loss}
\begin{align}
        \min_{\theta, A,B,q_\text{v},P, b} \text{    }& \mathbb{E}_\mathcal{D} \sum_{j=k}^{k+T_f} \left(\Vert \psi_\theta(y_{j}, d_{{j}})-\psi_{j|k}\Vert_2^2 +\Vert c_{j}-\
        \hat{c}_{j|k}(\psi_{j|k})\Vert_2^2 \right) \label{eq:obj1}\\ 
        \text{s.t. } 
        &\psi_{j+1|k} = A\psi_{j|k}+Bu_{j} \label{eq:linear propogation}\\ 
        &\psi_{k|k} = \psi_\theta(y_{k}, d_{{k}}) \label{eq:initial condition}
\end{align}
\end{subequations}
where 
$\mathcal{D}:= \{[y_k, u_k, d_k, c_{k}]^i_{0:T_f}\}_{i=1:N}$ denotes the sampled dataset; $\mathbb{E}_\mathcal{D}(\cdot)$ denotes calculating the expectation of variables over the data distribution of $\mathcal{D}$. 
In objective function \eqref{eq:obj1}, the first term is introduced to regulate the Koopman operator to learn an observable space where the dynamics are described using a linear form. The second term is introduced to minimize the reconstruction error of the stage cost, which is related to the precision of the model.

In our implementation, we solve the optimization problem described in \eqref{eq:loss} using a stochastic gradient descent algorithm called ADAM \cite{kingma_adam_2017}. The pseudocode for training the DIOKO model is shown in Algorithm~\ref{algo:DIOKO} below.
\begin{algorithm}
   \caption{Deep Input-Output Koopman Operator}
   \label{algo:DIOKO}
\begin{algorithmic}
    \REQUIRE
    Dataset $\mathcal{D}$, learning rate $\delta$, the maximum episode number $N$ and the batch number $M$
   \FOR{$i=1$ to $N$}
   \FOR{$j=1$ to $M$}
   \STATE Sample a batch of data from $\mathcal{D}$ and update $\theta, A,B,q_\text{v},P, b$ by using gradient descent with respect to the objective function \eqref{eq:loss}.
   \ENDFOR
   \STATE Evaluate the modeling performance.
   \ENDFOR
\end{algorithmic}
\end{algorithm}

\begin{remark}
\textcolor{black}{The proposed approach does not rely on an inverse mapping of observables to the states. Another advantage of the proposed method is its ability to directly predict state costs using the resulting Koopman operators, thereby eliminating the need for reconstructing the entire state vector of a process.}
In simulations, we have attempted to parameterize the output function in \eqref{eq:output function} as an affine function, that is, to use an output function that excludes the quadratic term, which is the first term on the right-hand-side of \eqref{eq:output function}. This adjustment yielded modeling performance comparable to the current formulation. However, when it comes to the control part considered in Section \ref{chapter:EMPC}, the controller consistently produces constant actions, either at the lower bound or the upper bound. This is because solutions to optimization problems with affine objective functions always reside at the boundary. It demonstrates that parameterization forms with similar modeling performance may significantly differ in terms of the well-posedness of the resulting control problem.
\end{remark}

\begin{remark}
\textcolor{black}{The proposed DIOKO can accurately predict the future economic cost based on the current output measurements and known disturbances, and it does not need access to the full-state information or future disturbances, which showcases its inherent robustness to information losses and unknown disturbances. To further deal with uncertainties induced by process and observation noise, one may encode the distribution of observables by using probabilistic neural networks as discussed in \cite{han2021desko}, which is left for future research. }
\end{remark}


\subsection{Data-driven convex EMPC}
\label{chapter:EMPC}
Based on the established DIOKO model, we develop a data-driven EMPC method for the economic and efficient operation of the wastewater treatment process. To achieve this objective, the economic predictive controller formulates and solves the following optimization problem:
\begin{subequations}\label{eq:economic MPC}
\begin{align}
    \min_{u_{k:k+T_f-1}|k} \sum_{j=k}^{k+T_f} \big(\psi_{j|k}^\text{T}& Q \psi_{j|k} +  P\psi_{j|k}+ b\big) + \sum_{j=k}^{k+T_f-1}\left(\Delta u_{j}^\text{T} R \Delta u_{j}\right)\label{eq:economic MPC1}\\
    \text{s.t. } ~\psi_{k+j+1|k} &= A \psi_{k+j|k} + B u_{k+j|k} \\
    \psi_{k|k} &= \psi_\theta(y_{k}, d_{{k}})\\
    \Delta u_{j|k} &= u_{j+1|k} - u_{j|k}\\
    u_{k+j}&\in \mathbb{U}
\end{align}
\end{subequations}
where $R$ is a positive (or semi-positive) definite weighting matrix, and the term weighted by $R$ is introduced to smooth the changes in control input between every two consecutive time steps. At each time instant $k$, the EMPC problem in \eqref{eq:economic MPC} is solved using a quadratic programming solver, and we obtain the optimal controls $u^*_{k:k+T_f-1|k}$. Then, only the first control input $u^*_{k|k}$ is applied to the process.

\begin{remark}
Note that the control scheme described in \eqref{eq:economic MPC} relies only on partial measurements of the full-state to generate optimal control inputs, whereas the existing economic model predictive control schemes for wastewater treatment process operations were developed based on the assumption of full access to the full-state $x_k$ \cite{ellis2014tutorial, Yin_8, zhang2019distributed}. Therefore, the method proposed in this work is more practical than the existing methods, considering the fact that obtaining full-state measurements for the WWTP can be challenging or at least expensive.
\end{remark}

\begin{remark}
The EMPC scheme presented in \eqref{eq:economic MPC} has a quadratic objective function and affine constraints. Therefore, its optimization can be efficiently addressed using quadratic programming. In contrast, most of the existing EMPC schemes for nonlinear systems rely on solving computationally expensive nonlinear optimizations. The proposed scheme is well-suited for the real-time control of the WWTP, and also holds the promise for economic and efficient operations of other high-dimensional nonlinear industrial processes. 
\end{remark}

\begin{remark}
\textcolor{black}{DIOKO focuses on building an input-output Koopman model instead of a dynamic model to characterize the dynamics of the full state of the process. This ensures the applicability of this method to even more complex water treatment facilities, since the control objective is expected to remain consistent, and the scale of the resulting DIOKO model will not increase significantly even with a much larger number of state variables for a complex facility. 
In terms of generalizability, While the emphasis of this research is placed on modeling and control of the water treatment process, the proposed method can be extended to address general nonlinear systems whose dynamics are described by \eqref{eq:process model:f}. For other nonlinear processes of which the dynamics are governed by this nonlinear state-space model, with an economic control objective, the proposed method holds the potential for establishing an input-output Koopman model and an efficient EMPC scheme.}
\end{remark}

\begin{figure}[!htb]
    \centering
        \subfigure[Trajectory of validation loss]{\label{fig:modeling performance-val}
        \includegraphics[width=0.47\textwidth]{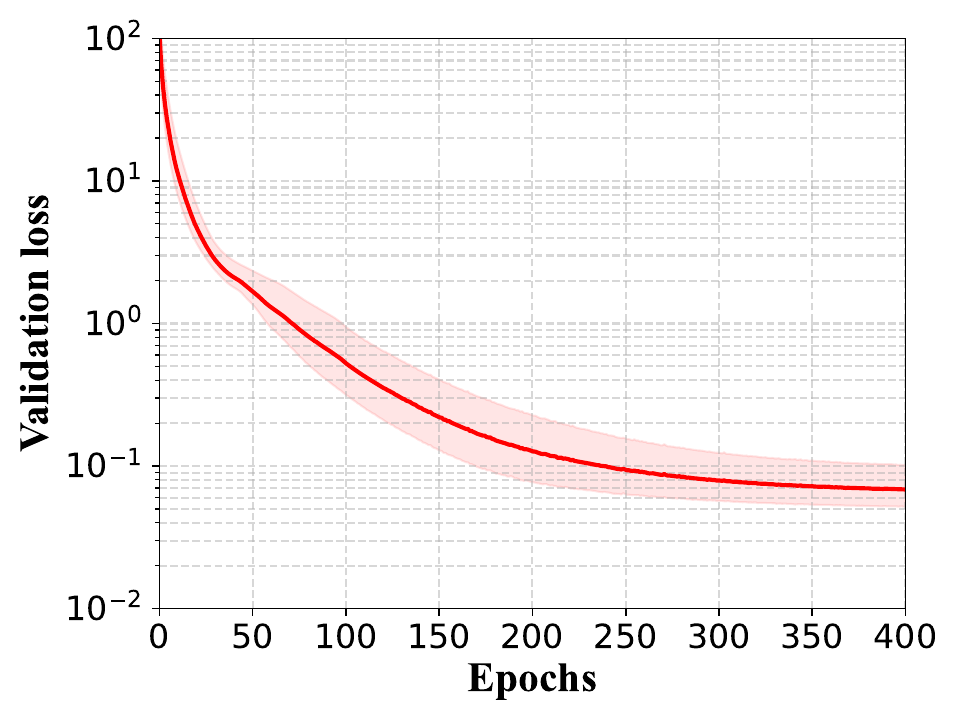}}
    \subfigure[Trajectory of test loss]{\label{fig:modeling performance-test}
        \includegraphics[width=0.47\textwidth]{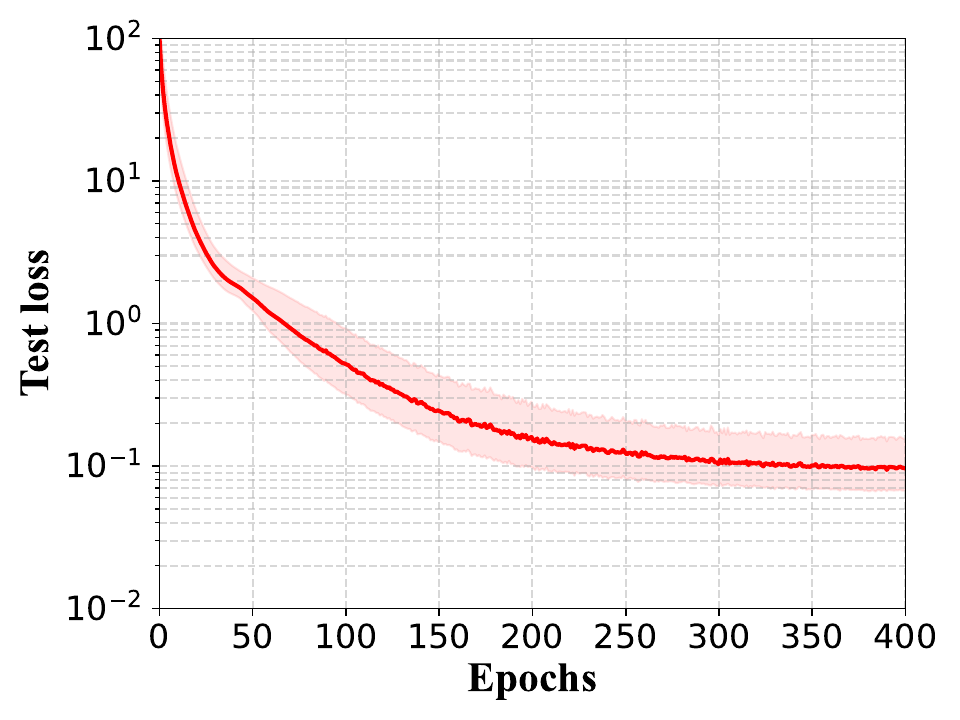}}
    \caption{Cumulative prediction error of the proposed DIOKO modeling approach on the validation data set and the test data set. The Y-axis indicates the cumulative mean-squared prediction error in log space over 16 instants, and the X-axis indicates the training epochs. The shaded region shows the confidence interval (one standard deviation) over 10 random initializations.}
    \label{fig:modeling performance}
\end{figure}



\begin{table}[!htb]
\centering
\caption{Initial Condition of the States in the Secondary Settler}\vspace{2mm}
\label{table:initial condition of SS_1}\renewcommand\arraystretch{1.27}
\begin{tabular}{p{1.3cm}|p{1.2cm}p{1.2cm}p{1.2cm}p{1.7cm}}
\hline
Layer index $i$& $X_\text{i}$ & $S_\text{I,i}$ & $S_\text{S,i}$& $S_\text{O,i}$ \\
\hline

1&6399.44&30.0&0.808&2.0\\
2&356.29&30.0&0.808&2.0\\
3&356.29&30.0&0.808&2.0\\
4&356.29&30.0&0.808&2.0\\
5&356.29&30.0&0.808&2.0\\
6&356.29&30.0&0.808&2.0\\
7&69.00&30.0&0.808&2.0\\
8&29.55&30.0&0.808&2.0\\
9&18.12&30.0&0.808&2.0\\
10&12.50&30.0&0.808&2.0\\
units&$\text{gCOD}/\text{m}^3$&$\text{gCOD}/\text{m}^3$&$\text{gCOD}/\text{m}^3$&$\text{g(-COD)}/\text{m}^3$\\
\hline
\end{tabular}
\end{table}

\begin{table}[!htb]
\centering
\caption{Initial Condition of the States in the Secondary Settler}\vspace{2mm}
\label{table:initial condition of SS_2}\renewcommand\arraystretch{1.27}
\begin{tabular}{p{1.3cm}|p{1.2cm}p{1.2cm}p{1.2cm}p{1.7cm}}
\hline
Layer index $i$& $S_\text{NO,i}$ & $S_\text{NH,i}$ &$S_\text{ND,i}$& $S_\text{ALK,i}$ \\
\hline

1&13.52&0.67&0.66&3.83\\
2&13.52&0.67&0.66&3.83\\
3&13.52&0.67&0.66&3.83\\
4&13.52&0.67&0.66&3.83\\
5&13.52&0.67&0.66&3.83\\
6&13.52&0.67&0.66&3.83\\
7&13.52&0.67&0.66&3.83\\
8&13.52&0.67&0.66&3.83\\
9&13.52&0.67&0.66&3.83\\
10&13.52&0.67&0.66&3.83\\
units&$\text{gN}/\text{m}^3$&$\text{gN}/\text{m}^3$&$\text{gN}/\text{m}^3$&$\text{mol}/\text{m}^3$\\
\hline
\end{tabular}
\end{table}

\begin{table}[!htb]
\centering
\caption{Hyper-parameters of DIOKO}\vspace{2mm}
\label{table:hyperparameters}\renewcommand\arraystretch{1.27}
\begin{tabular}{l|c }
\hline
Hyper-parameters&Value\\
\hline
Size of data set $\mathcal{D}$& $10^5$\\
Batch Size & $128$\\
Epochs & $400$\\
Learning rate & $10^{-3}$\\
Prediction horizon $T_\text{f}$ & $30$ \\
Structure of $\psi_\theta(\cdot)$ &$(128,128)$\\
Activation function & ELU\\
Dimension of observables & $60$\\
$l_2$ norm regularization coefficient & $0.1$\\
\hline
\end{tabular}
\end{table}

\begin{table*}
\centering
\caption{The summary of the four key performance indices based on the evaluated control methods}\vspace{2mm}
\label{table:performance summary}\renewcommand\arraystretch{1.27}
\begin{tabular}{p{2cm}|p{3cm}|p{2cm}p{2cm}p{2cm}p{2.4cm}}
\hline
Weather&Methods& Stage costs & EQ & OCI&Computation time\\
\hline
\multirow{5}{4em}{Dry}&DIOKO EMPC&$\mathbf{1.295}\times 10^7$&$\mathbf{5.926}\times 10^6$&$2.341\times 10^7$& ${0.0339}$ s\\
&EMPC \cite{Yin_8}&$1.582\times 10^7$&$8.893\times 10^6$&$2.310\times 10^7$& 334.5 s\\
&MPC &$1.789\times 10^7$&$11.22\times 10^6$&$\mathbf{2.223}\times 10^7$& 17.36 s\\
&DIOKO-dry &$1.386\times 10^7$&$6.753\times 10^6$&$2.371\times 10^7$& 0.0341 s\\
&\textcolor{black}{SAC \cite{haarnoja2018soft}}&\textcolor{black}{${1.345}\times 10^7$}&\textcolor{black}{${6.346}\times 10^6$}&\textcolor{black}{$2.366\times 10^7$}& \textcolor{black}{$\mathbf{0.0046}$ s}\\
\hline
\multirow{5}{4em}{Rainy}&DIOKO EMPC&$\mathbf{1.219}\times 10^7$&$\mathbf{6.519}\times 10^6$&$1.893\times 10^7$& 0.0361 s\\
&EMPC \cite{Yin_8}&$1.592\times 10^7$&$10.15\times 10^6$&$1.920\times 10^7$& 249.5 s\\
&MPC &$1.841\times 10^7$&$12.93\times 10^6$&$\mathbf{1.825}\times 10^7$& 16.03 s\\
&DIOKO-dry &$1.369\times 10^7$&$7.766\times 10^6$&$1.973\times 10^7$&$ {0.0353}$ s\\
&\textcolor{black}{SAC \cite{haarnoja2018soft}}&\textcolor{black}{${1.331}\times 10^7$}&\textcolor{black}{${7.355}\times 10^6$}&\textcolor{black}{$1.984\times 10^7$}& \textcolor{black}{$\mathbf{0.0041}$ s}\\
\hline
\multirow{5}{4em}{Stormy}&DIOKO EMPC&$\mathbf{1.343}\times 10^7$&$\mathbf{6.036}\times 10^6$&$2.465\times 10^7$& 0.0368 s\\
&EMPC \cite{Yin_8}&$1.722\times 10^7$&$9.822\times 10^6$&$2.468\times 10^7$& 206.9 s\\
&MPC &$1.928\times 10^7$&$12.11\times 10^6$&$\mathbf{2.391}\times 10^7$& 17.14 s\\
&DIOKO-dry &$1.458\times 10^7$&$7.022\times 10^6$&$2.521\times 10^7$& $\mathbf{0.0351}$ s\\
&\textcolor{black}{SAC \cite{haarnoja2018soft}}&\textcolor{black}{${1.441}\times 10^7$}&\textcolor{black}{${6.797}\times 10^6$}&\textcolor{black}{$2.537\times 10^7$}& \textcolor{black}{$\mathbf{0.0040}$ s}\\
\hline
\end{tabular}
\end{table*}

\begin{figure}[!htb]
    \centering
    \subfigure[Trajectories of the stage cost of the proposed method and baseline methods.]{
        \label{fig:stage cost-dry}
        \includegraphics[width=0.47\textwidth]{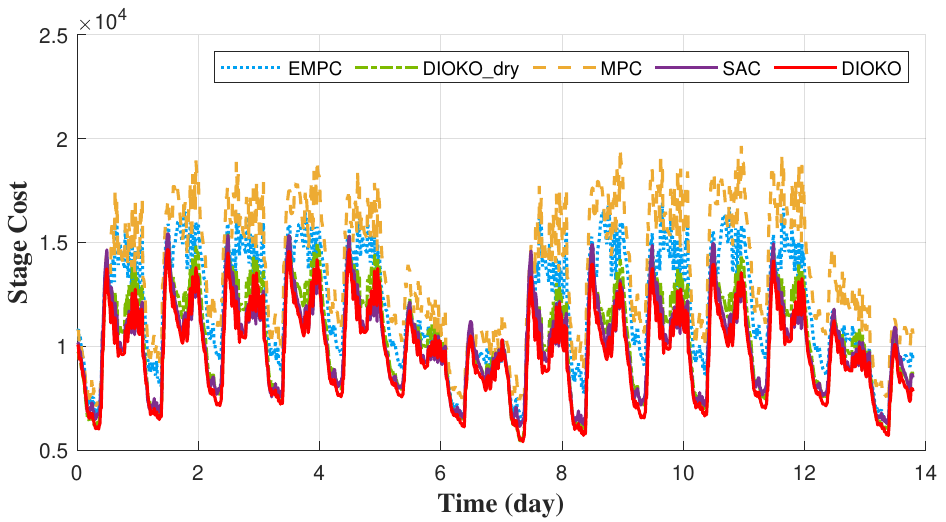}}
    \subfigure[Trajectories of the EQ of the proposed method and baseline methods.]{
        \label{fig:eq-dry}
        \includegraphics[width=0.47\textwidth]{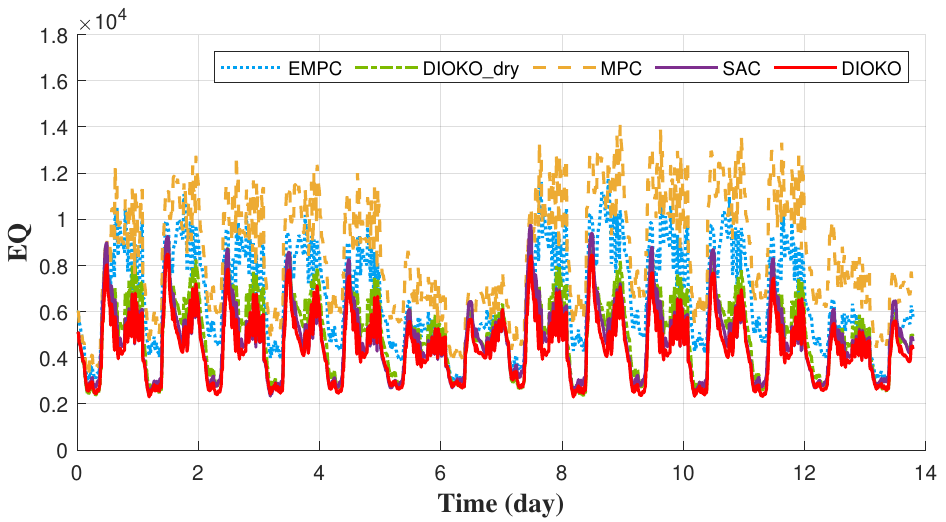}}
    \subfigure[Trajectories of the OCI of the proposed method and baseline methods.]{
        \label{fig:oci-dry}
        \includegraphics[width=0.47\textwidth]{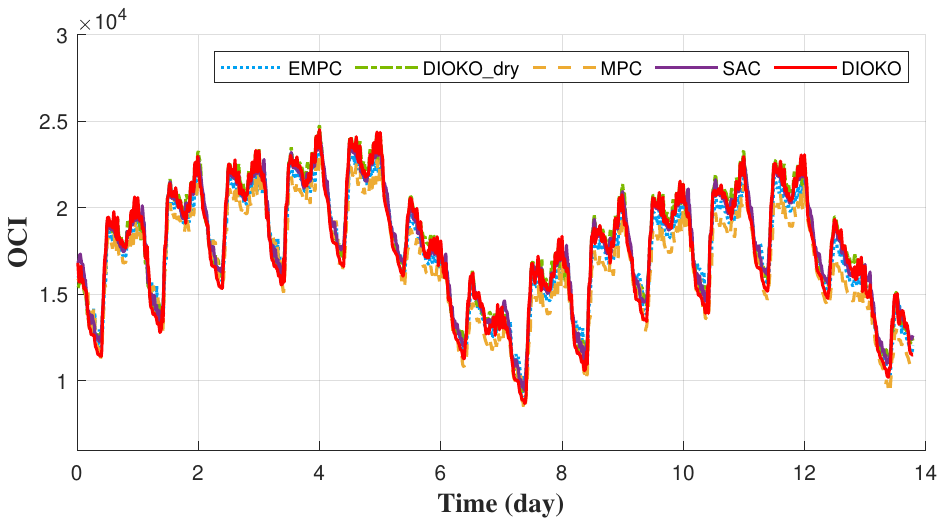}}
    \caption{Control performance comparison between the proposed DIOKO-based EMPC and baselines under the dry weather conditions. The trajectories of the overall economic stage cost ($c_k$), EQ, and OCI for each of the methods are presented.}
    \label{fig:control dry}
\end{figure}

\section{Simulation results}

In this section, we apply the proposed Koopman-based modeling and economic predictive control method to the WWTP of which the dynamics are characterized by the BSM1. The modeling accuracy, control performance, and computational efficiency of our proposed DIOKO framework are assessed. We also compare the control performance of the proposed method with two representative baselines, including the EMPC designed for WWTP based on a nonlinear first-principles model \cite{Yin_8} and the conventional nonlinear tracking MPC.

\subsection{Simulation settings}\label{subsec:simulation settings}

We consider three different weather conditions, including dry, rainy, and stormy weather conditions, and apply the proposed method to the WWTP operated under these different conditions, for which the data files can be found on the International Water Association website\footnote{\url{https://iwa-network.org/}}.

Tables~\ref{table:initial condition of BR}, \ref{table:initial condition of SS_1} and \ref{table:initial condition of SS_2} present the initial conditions of the state variables of the activated sludge reactor and the secondary settler, respectively. The initial conditions are determined through an open-loop steady-state simulation of the process operation, with constant control inputs and known disturbances applied to the process for 14 days under dry weather conditions, in the same manner as outlined in \cite{o2011model, Yin_8}. The corresponding steady-state values are used as the initial condition of the process operation, as shown in the two tables. The plant is simulated for 14 days to evaluate the performance of the controllers with a sampling interval of $T = 15$ minutes.

The weighting matrix is designed as 
\begin{equation}
    R = \left[ \begin{matrix}
        1.2 \times 10^{-8} & 0 \\
        0 & 1.77733\times 10^{-5}
    \end{matrix}\right]
\end{equation}
and the control inputs are constrained to the range of $0\leq\text{Q}_\text{a}\leq 92230~\text{m}^3/\text{day}$ and $0\leq\text{K}_\text{L}\text{a}_5\leq 240~\text{day}^{-1}$. The weights in \eqref{eq:stage cost} are set to be $w_\text{EQ}=1$ and $w_\text{OCI}=0.3$. In the design of the MPC baseline, the control objective is to maintain the nitrate concentration of the second anoxic reactor, $S_\text{NO,2}$, and the dissolved oxygen concentration of the last aerobic reactor, $S_\text{O,5}$, at the desired set-points $\bar{S}_\text{NO,2}= 1 \text{gN}/\text{m}^3$ and $\bar{S}_\text{O,5} = 2 \text{gCOD}/\text{m}^3$, respectively.

One potential challenge of implementing the proposed DIOKO modeling method in a real-world plant is the collection of an informative data set. The input signal needs to excite the system while maintaining operational safety. Knowledge from the domain expert is essential in designing such input signals. In our test, the data set is collected by applying a series of inputs, $u_k = \bar{u} + \epsilon_k$ to the system under different weather conditions, where $\bar{u}$ is generated uniformly from the action space $\mathbb{U}$ and updated every $20$ steps, and noise $\epsilon_k$ drawn from the normal distribution $\mathcal{N}(0, 0.05\bar{u})$ and updated at every instant. This random input sequence maintains a certain level of randomness while preventing the occurrence of excessively significant changes. At the initial instant, one of the three weather conditions is randomly chosen and the process is simulated for a duration of 14 days. In total, $10^5$ data samples are collected. Among them, the first $8\times10^4$ samples are used for training, another $10^4$ samples constitute the validation set, and the remaining $10^4$ samples constitute the test set. Hyperparameters of DIOKO are shown in Table~\ref{table:hyperparameters}.

For a comparison purpose, we adopt the tracking model predictive control (MPC)  and the economic model predictive control (EMPC) \cite{Yin_8} as the baselines. The tracking MPC solves the optimal control problem for the WWTP by tracking predefined set-points. The EMPC approach solves the same optimal control problem for the WWTP in the sense that the economic stage cost is identical to the one considered in the proposed method. Meanwhile, the implementation of the MPC and EMPC designs based on the nonlinear first-principles process model requires full-state information $x_k$, while the proposed DIOKO-based EMPC only requires partial state measurements $y_k$ that are directly related to the economic stage cost $c_k$. Therefore, this comparison inherently favors MPC and EMPC in terms of information availability. Furthermore, we have considered a model-free reinforcement learning algorithm -- soft actor-critic (SAC) \cite{haarnoja2018soft}, as a learning-based control baseline. In the training phase, SAC relies on the identical economic stage cost as EMPC.

\subsection{Modeling evaluation}
First, we evaluate the performance of the proposed DIOKO modeling method. In total, 10 DIOKO models with random parameter initializations are trained. Each model undergoes 400 training epochs, where, in each epoch, the algorithm takes a batch of 128 data points from the data set and updates the parameters until the data set has been completely traversed. Figures~\ref{fig:modeling performance-val} and \ref{fig:modeling performance-test} present the $l_2$ norms of the prediction losses for the validation and test data sets for each epoch, respectively. It can be seen from Figure~\ref{fig:modeling performance} that both the validation and the test losses converge to a magnitude of $10^{-1}$ as training proceeds. In addition, the shadowed region in Figure~\ref{fig:modeling performance} indicates that this convergence is not affected by parameter initializations.
\begin{figure}[!htb]
    \centering
    \subfigure[Trajectories of the stage cost of the proposed method and baseline methods.]{
        \label{fig:stage cost-rain}
        \includegraphics[width=0.47\textwidth]{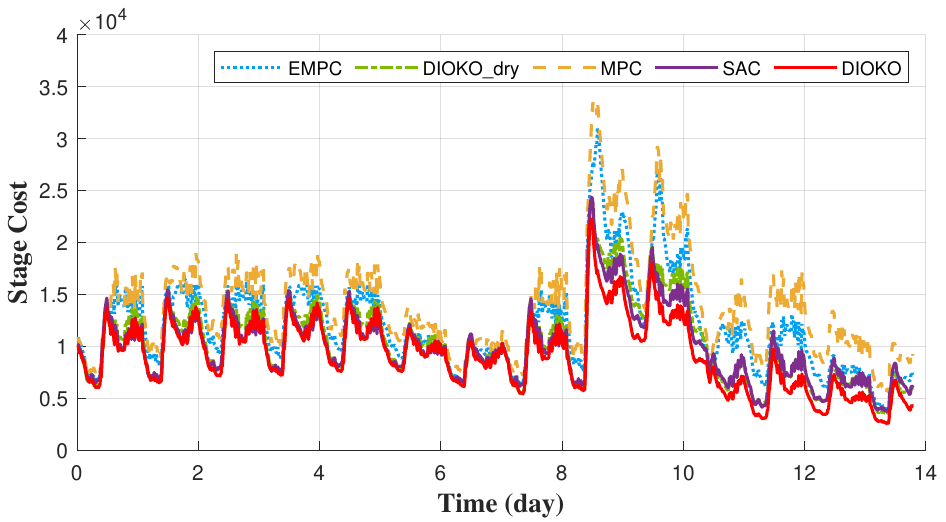}}
    \subfigure[Trajectories of the EQ of the proposed method and baseline methods.]{
        \label{fig:eq-rain}
        \includegraphics[width=0.47\textwidth]{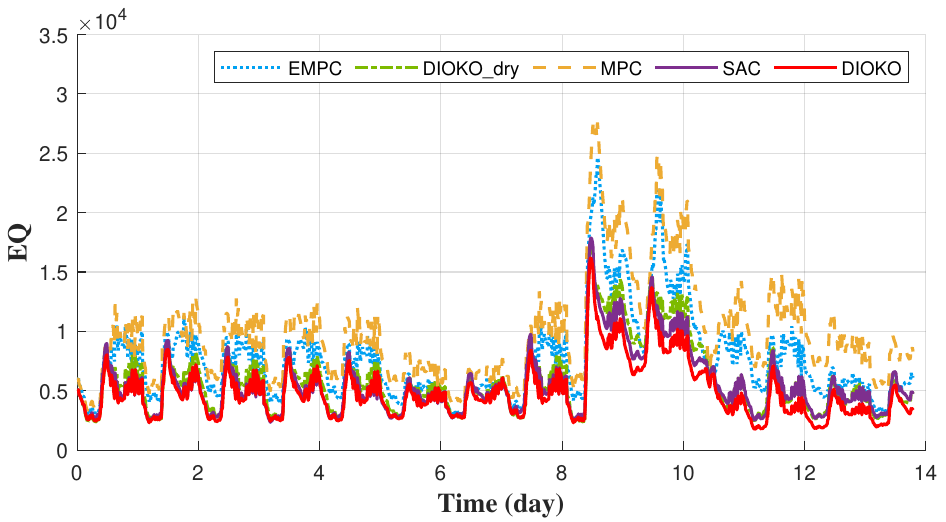}}
    \subfigure[Trajectories of the OCI of the proposed method and baseline methods.]{
        \label{fig:oci-rain}
        \includegraphics[width=0.47\textwidth]{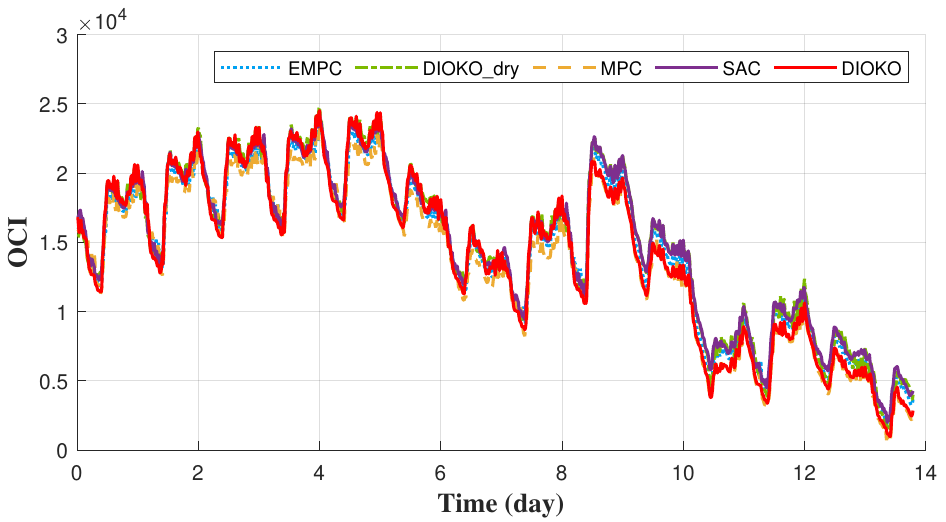}}
    \caption{Control performance comparison between the proposed DIOKO-based EMPC and baselines under the rainy weather condition. The trajectories of the overall economic stage cost ($c_k$), EQ, and OCI for each of the methods are presented.}
    \label{fig:control rain}
\end{figure}

\subsection{Control evaluation}

\begin{figure}[!htb]
    \centering
    \subfigure[Trajectories of the stage cost of the proposed method and baseline methods.]{
        \label{fig: stage cost-strm}
        \includegraphics[width=0.47\textwidth]{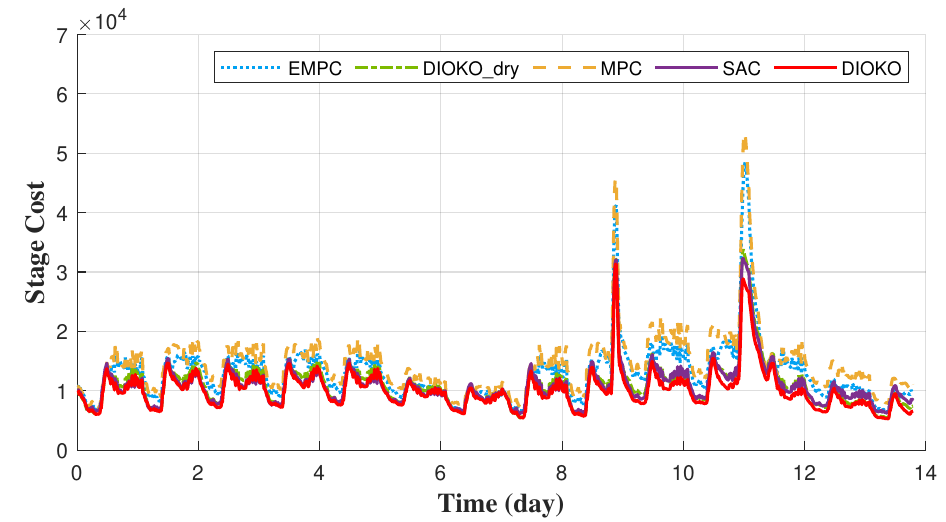}}
    \subfigure[Trajectories of the EQ of the proposed method and baseline methods.]{
        \label{fig:eq-strm}
        \includegraphics[width=0.47\textwidth]{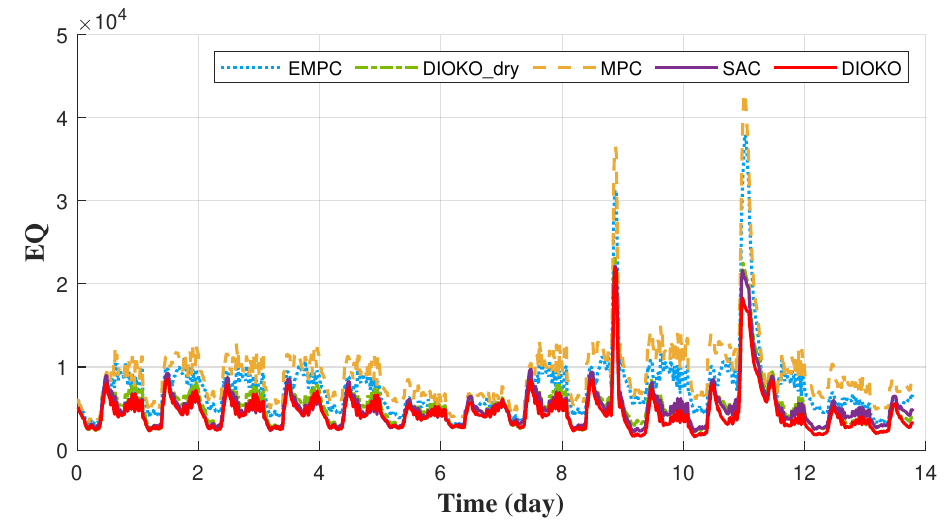}}
    \subfigure[Trajectories of the OCI of the proposed method and baseline methods.]{
        \label{fig:oci-strm}
        \includegraphics[width=0.47\textwidth]{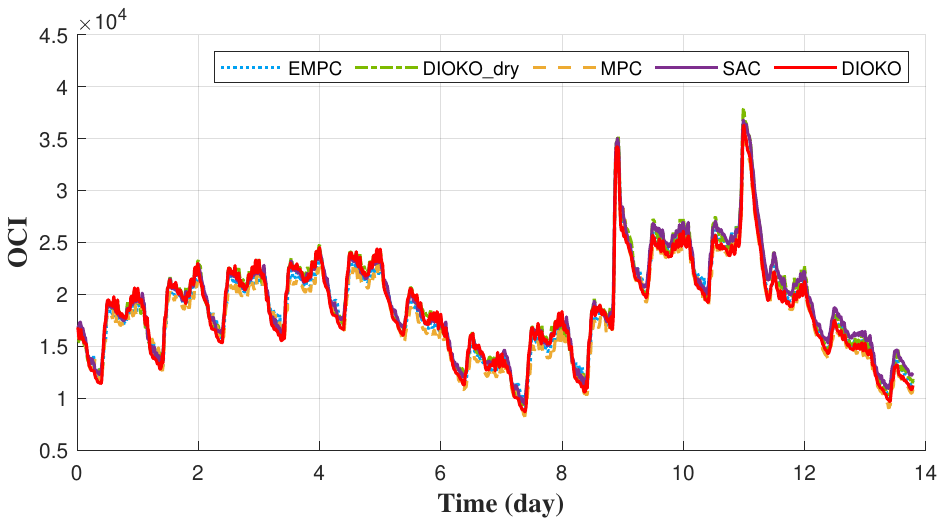}}
    \caption{Control performance comparison between the proposed DIOKO-based EMPC and baselines under the stormy weather condition. The trajectories of the overall economic stage cost ($c_k$), EQ, and OCI for each of the methods are presented.}
    \label{fig:control strm}
\end{figure}

In this subsection, we conduct a comparative analysis of the control performance between the proposed DIOKO EMPC method, set-point tracking nonlinear MPC method and the EMPC method in \cite{Yin_8}. The proposed method and the baselines are evaluated under three weather conditions, including dry weather, rainy weather, and stormy weather. The corresponding results are presented in Figures~\ref{fig:control dry}, \ref{fig:control rain} and \ref{fig:control strm}, respectively, where the real-time overall economic stage cost, effluent quality (EQ), and overall cost index (OCI) of the control schemes are depicted in separate sub-figures. 
These performance indices of the compared methods under different weather conditions are summarized in Table~\ref{table:performance summary}. The bold numbers indicate the best performance among the four approaches.

An examination of Table~\ref{table:performance summary} reveals that the proposed method provides the lowest economic stage cost and the best effluent quality under all three weather conditions, as compared to the baselines.
In terms of the overall economic stage cost, which is computed based on OCI and EQ following (\ref{eq:stage cost}), the performance provided by the proposed DIOKO EMPC is improved by $18.1\% \sim 33.8\%$ compared to the model-based baseline methods, subject to different weather conditions. In terms of the effluent quality, the performance obtained by the proposed DIOKO-based EMPC is improved by $33.4\% \sim 50.2\%$ as compared to the model-based baseline methods. This confirms that the proposed method is able to enhance the overall operational performance of the WWTP, particularly the effluent quality, under various weather conditions. In comparison to SAC, DIOKO achieves a lower overall stage cost and superior effluent quality with significantly less data. The training of SAC uses $10^6$ data points for convergence, which is 10 times the size of the data set for DIOKO.

\subsection{Computational efficiency}
One significant advantage of the proposed DIOKO-based approach is its capability to bypass the need to solve a complex nonlinear optimization problem. The proposed method formulates a quadratic programming problem that can be efficiently solved. This is confirmed by the simulation results obtained under various weather conditions as well. The average computation times per step for the proposed method and the baseline EMPC are presented in the last column of Table~\ref{table:performance summary}. Both of the approaches were executed on a workstation equipped with an i9-13900KF CPU. As shown in the table, DIOKO is able to solve the optimal control problem within 3 to 4 ms on average, which is approximately 5600 to 9800 times more efficient than the baseline EMPC based on the nonlinear process model, and 400 to 500 times more efficient than the MPC approach. From an online implementation point of view, SAC is the most computationally efficient method, which only takes $0.004$ seconds for each control step. SAC directly maps the state measurement to control input by using a neural network and eliminates the need for solving optimization in the predictive control methods.

\subsection{Robustness}

\begin{figure}[!htb]
    \centering
        \subfigure[Trajectory of validation loss]{\label{fig:modeling performance-val_noisy}
        \includegraphics[width=0.47\textwidth]{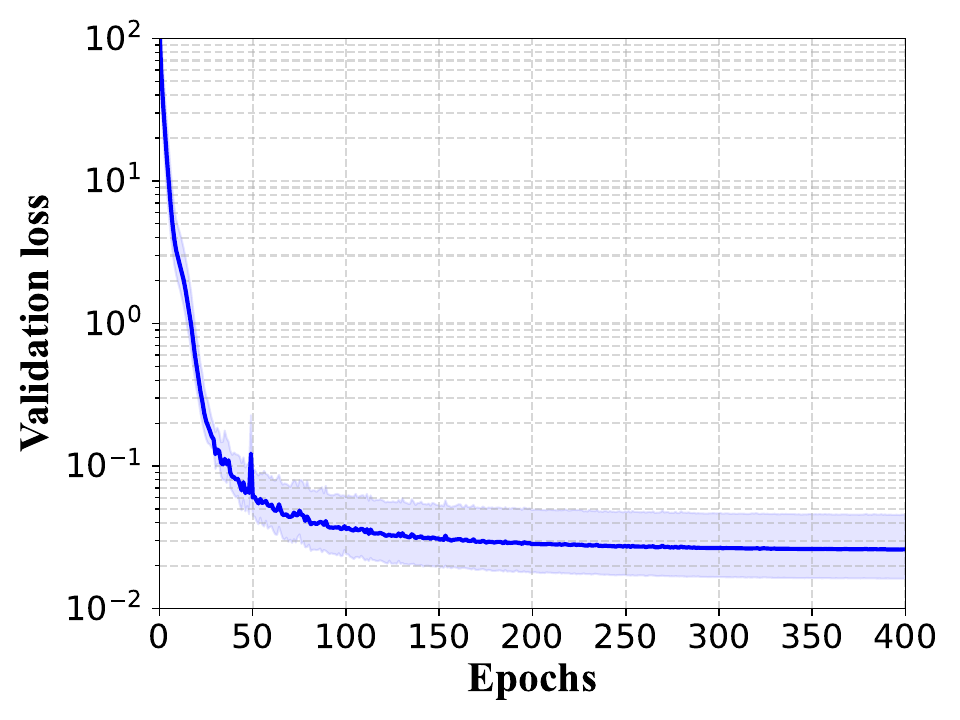}}
    \subfigure[Trajectory of test loss]{\label{fig:modeling performance-test_noisy}
        \includegraphics[width=0.47\textwidth]{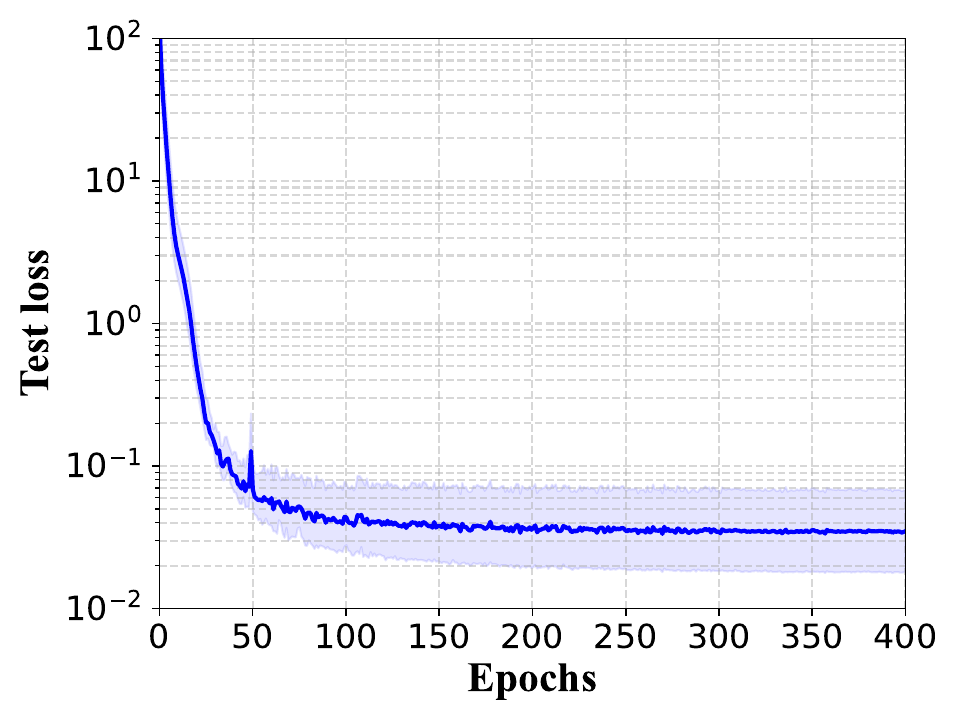}}
    \caption{Cumulative prediction error of the DIOKO-noisy on the validation data set and the test data set. The Y-axis indicates the cumulative mean-squared prediction error in log space over 16 instants, and the X-axis indicates the training epochs. The shaded region shows the confidence interval (one standard deviation) over 10 random initializations.}
    \label{fig: robust modeling performance}
\end{figure}

In practice, the system and the collected data set are usually subject to noise and uncertainties, Therefore, it is crucial to evaluate the robustness of the proposed DIOKO method against noise. 
To address this, we further train DIOKO with a noisy data set, denoted as \textit{DIOKO-noisy}, and evaluate its modeling performance and the control performance based on this model. Additionally, the DIOKO model trained on a clean data set is tested on a noisy data set. The process and measurements are corrupted by normally distributed noises, with zero mean and the scale of $0.001x_0$ and $0.0005\times x_0$, respectively. The hyperparameters of DIOKO-noisy are the same as described in Table~\ref{table:hyperparameters}. 
Figure~\ref{fig: robust modeling performance} illustrates the prediction performance of DIOKO-noisy on both the validation and test data sets. Notably, DIOKO trained on the noisy data set achieves a prediction accuracy similar to that observed on the clean data set. 

\begin{table}
\centering
\caption{\textcolor{black}{A summary of the control performance under noisy process operation}}\vspace{2mm}
\label{table:robust performance summary}\renewcommand\arraystretch{1.27}
\begin{tabular}{p{2.2cm}|p{1.6cm}p{1.6cm}p{1.6cm}}
\hline
Methods& Stage costs & EQ & OCI\\
\hline
\multicolumn{4}{l}{Dry weather}\\
\hline
\textcolor{black}{DIOKO EMPC} &\textcolor{black}{$1.340\times 10^7$}&\textcolor{black}{$6.454\times 10^6$}&\textcolor{black}{$2.317\times 10^7$}\\
\textcolor{black}{DIOKO-noisy}&\textcolor{black}{${1.368}\times 10^7$}&\textcolor{black}{${6.836}\times 10^6$}&\textcolor{black}{$2.281\times 10^7$}\\
\textcolor{black}{MPC} &\textcolor{black}{$1.777\times 10^7$}&\textcolor{black}{$11.19\times 10^6$}&\textcolor{black}{$2.195\times 10^7$}\\
\hline
\multicolumn{4}{l}{Rainy weather}\\
\hline
\textcolor{black}{DIOKO EMPC} &\textcolor{black}{$1.224\times 10^7$}&\textcolor{black}{$6.685\times 10^6$}&\textcolor{black}{$1.852\times 10^7$}\\
\textcolor{black}{DIOKO-noisy}&\textcolor{black}{${1.287}\times 10^7$}&\textcolor{black}{${7.096}\times 10^6$}&\textcolor{black}{$1.927\times 10^7$}\\
\textcolor{black}{MPC} &\textcolor{black}{$1.832\times 10^7$}&\textcolor{black}{$12.91\times 10^6$}&\textcolor{black}{\textbf{$1.801$}$\times 10^7$}\\
\hline
\multicolumn{4}{l}{Stormy weather}\\
\hline
\textcolor{black}{DIOKO EMPC} &\textcolor{black}{\textbf{$1.467$}$\times 10^7$}&\textcolor{black}{\textbf{$6.897$}$\times 10^6$}&\textcolor{black}{$2.590\times 10^7$}\\
\textcolor{black}{DIOKO-noisy}&\textcolor{black}{${1.470}\times 10^7$}&\textcolor{black}{${7.029}\times 10^6$}&\textcolor{black}{$2.558\times 10^7$}\\
\textcolor{black}{MPC} &\textcolor{black}{$1.938\times 10^7$}&\textcolor{black}{$12.06\times 10^6$}&\textcolor{black}{\textbf{$2.437$}$\times 10^7$}\\
\hline
\end{tabular}
\end{table}

Next we consider a noisy closed-loop water treatment process operation environment. In this context, unknown process disturbances and measurement noise are introduced to the BSM1 model as control is implemented online. The control performance of DIOKO, DIOKO-noisy, and MPC are evaluated and summarized in Table~\ref{table:robust performance summary}. DIOKO shows good robustness and provides the lowest economic cost under the three different weather. On the noisy BSM1, the economic cost achieved by DIOKO is still lower than the scores achieved by EMPC and MPC in the nominal case. 

\subsection{\textcolor{black}{Sensitivity}}

\begin{figure}[!htb]
    \centering
        \subfigure[\textcolor{black}{Sensitivity to learning rate}]{\label{fig:modeling performance_sensitivity_lr}
        \includegraphics[width=0.47\textwidth]{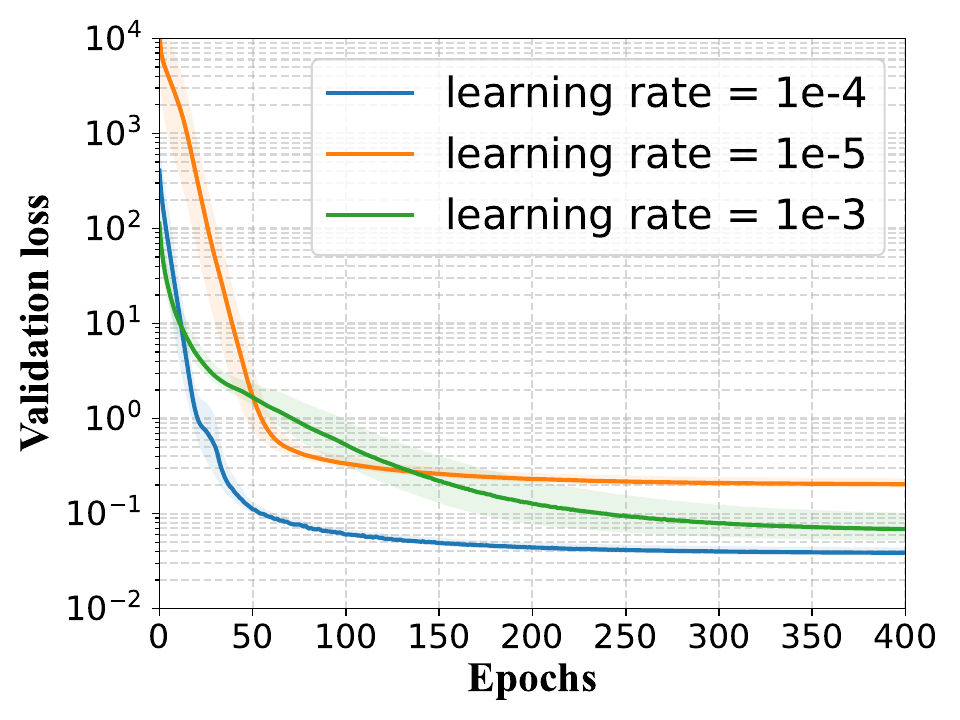}}
    \subfigure[\textcolor{black}{Sensitivity to latent dimension}]{\label{fig:modeling performance_sensitivity_latent_dim}
        \includegraphics[width=0.47\textwidth]{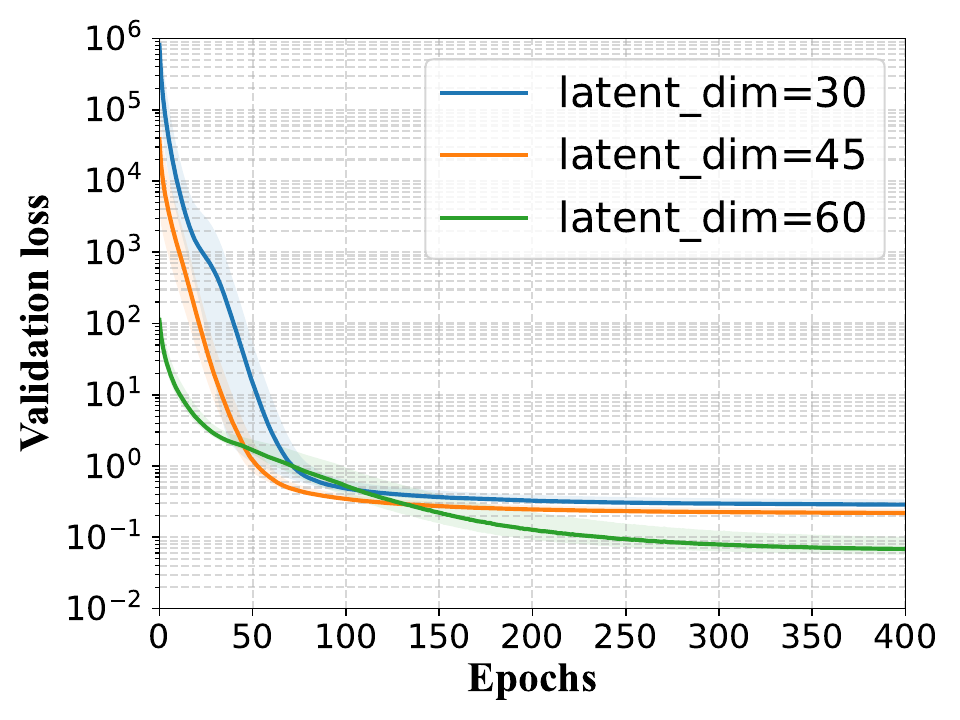}}
    \caption{\textcolor{black}{Cumulative prediction error of the DIOKO trained with different hyperparameters on the validation set. The validation loss }}
    \label{fig: modeling performance sensitivity}
\end{figure}

In this subsection, we analyze how the hyperparameters affect the performance of the DIOKO pipeline. Among the hyperparameters listed in Table~\ref{table:hyperparameters}, the learning rate and the dimension of the observables (i.e., latent dimension) can be particularly influential. Therefore, a sensitivity analysis is conducted on these two hyperparameters. Specifically, the learning rate is set at $10^{-5}$, $10^{-4}$ ,$10^{-3}$, or $10^{-2}$, and the latent dimension is chosen from 30, 45, or 60. While tuning one hyperparameter, the other ones remain the same as listed in Table~\ref{table:hyperparameters}. The modeling performance of the DIOKO models trained under different hyperparameters is shown in Figure~\ref{fig: modeling performance sensitivity}.

The model trained with the learning rate of $10^{-4}$ provides the smallest prediction error, while the learning rate of $10^{-2}$ causes divergence of the training process and is not included in Figure~\ref{fig: modeling performance sensitivity}. On the other hand, the latent dimension of $60$ leads to the smallest prediction error. Overall, the DIOKO performs consistently well across the different hyperparameter settings, and this indicates that the proposed method in general requires only limited effort in hyperparameter tuning. 

\subsection{Generalization}

For a learning-based modeling and control framework, it is also important to verify its generalizability, i.e., to examine its performance on previously unencountered data. To address this, we leverage the DIOKO model established based on a data set collected under the dry weather condition to develop a control scheme, which is referred to as DIOKO-dry. We evaluate the control performance of this scheme for the rainy and stormy weather conditions. The hyperparameters of the DIOKO-dry and the size of the data set remain the same as the settings introduced in Section~\ref{subsec:simulation settings}. The results are shown in Figure~\ref{fig:control dry}, \ref{fig:control rain}, and \ref{fig:control strm}, and summarized in Table~\ref{table:performance summary}. 

By examining the table, it shows that DIOKO-dry obtains clear improvement compared to the baselines in all three weather conditions. To be specific, under the dry weather condition, the overall economic stage cost of DIOKO-dry is improved by $12.3\% \sim 22.5\%$ as compared to the baseline methods. In terms of the effluent quality, the performance obtained by DIOKO-dry is improved by $24.1\% \sim  39.8\%$ as compared to the baseline methods. Furthermore, in the previously unencountered rainy and stormy weather conditions, the overall economic stage cost is improved by $14.0\% \sim 25.6\%$, and the effluent quality is improved by $23.5\% \sim  42.0\%$. In conclusion, the DIOKO trained using the data set for the dry weather condition exhibits adequate ability to generalize to other weather conditions and provide superior economic performance, in comparison to the baseline methods. 

A comparison between DIOKO-dry and the model established based on the proposed approach, referred to as DIOKO EMPC, which is trained using data sets for all three weather conditions, evidences the benefits of using a diverse data set for learning-based Koopman modeling within the proposed framework. The proposed DIOKO EMPC outperforms the DIOKO-dry in terms of the stage costs, the effluent quality, and the overall process operation performance, under each of the three weather conditions, including the dry weather condition. This indicates that the use of a diverse data set may help enhance the performance of the learned model and the resulting control scheme.

\section{Concluding remarks}

In this study, we proposed a learning-based input-output Koopman modeling and economic predictive control approach for the used water treatment process. 
Within the Koopman operator framework, a data-driven model, referred to as the DIOKO model, which can predict the overall economic cost of the WWTP operation was established based on the input data and available output measurements. By using the DIOKO model, a convex economic predictive control scheme was formulated. The proposed method was applied to the WWTP under three weather conditions, and extensive comparisons were made among the proposed method, the existing EMPC design for the WWTP, and a representative nonlinear MPC design. Good predictability was achieved without the need to incorporate first-principles knowledge into the model or require full-state measurements. As compared to the conventional control approach, the proposed framework has improved the overall economic operational performance which balances the effluent quality and the overall cost index. Additionally, our proposed method demonstrated a remarkable over 5600-fold improvement in computational efficiency as compared to the existing non-convex EMPC design for the wastewater treatment process. 

\textcolor{black}{We also identify some important topics that need to be explored in future research. First, although the DIOKO model is shown to be robust to noise, the method does not formally take into account the unknown disturbances. In this context, the development of probabilistic input-output Koopman modeling would be interesting. Second, in case of scenarios when first-principles knowledge is available yet only limited data is accessible, it is worthwhile to investigate the integration of these two information sources for hybrid Koopman modeling and control.}












\end{document}